\documentclass[natbib,twocolumn]{aastex631}

\usepackage{graphicx}	% Including figure files
\usepackage{amsmath}	% Advanced maths commands
\usepackage{amssymb}	% Extra maths symbols
\usepackage{bm}		% Bold maths symbols, including upright Greek
\usepackage{upgreek}	% upright Greek
\usepackage{xcolor}	% more colors

\shorttitle{Subtle and Spectacular Debris Disks}
\shortauthors{Farihi et al.}

\begin{document}

\title{Subtle and Spectacular: Diverse White Dwarf Debris Disks Revealed by {\em JWST}}

\correspondingauthor{J.~Farihi}
\email{j.farihi@ucl.ac.uk}

\author[0000-0003-1748-602X]{J.~Farihi}
\affiliation{Department of Physics and Astronomy, University College London, London WC1E 6BT, UK}

\author[0000-0002-3532-5580]{K.~Y.~L.~Su}
\affiliation{Space Science Institute, Boulder CO 80301, USA}

\author[0000-0001-9834-7579]{C.~Melis}
\affiliation{Astronomy \& Astrophysics Department, University of California, San Diego CA 92093-0424, USA}

\author[0000-0003-0214-609X]{S.~J.~Kenyon}
\affiliation{Smithsonian Astrophysical Observatory, 60 Garden Street, Cambridge, MA 02138, USA}

\author[0000-0001-6515-9854]{A.~Swan}
\affiliation{Department of Physics, University of Warwick, Coventry CV4 7AL, UK}

\author[0000-0003-3786-3486]{S.~Redfield}
\affiliation{Astronomy Department and Van Vleck Observatory, Wesleyan University, Middletown CT 06459, USA}

\author[0000-0001-9064-5598]{M.~C.~Wyatt}
\affiliation{Institute of Astronomy, University of Cambridge, Madingley Road, Cambridge CB3 0HA, UK}

\author[0000-0002-1783-8817]{J.~H.~Debes}
\affiliation{Space Telescope Science Institute, Baltimore, MD 21218, USA}

\begin{abstract}

This letter reports 12 novel spectroscopic detections of warm circumstellar dust orbiting polluted white dwarfs using {\em JWST} MIRI.  The disks span two orders of magnitude in fractional infrared brightness and more than double the number of white dwarf dust spectra available for mineralogical study.  Among the highlights are: i) the two most subtle infrared excesses yet detected, ii) the strongest silicate emission features known for any debris disk orbiting any main-sequence or white dwarf star, iii) one disk with a thermal continuum but no silicate emission, and iv) three sources with likely spectral signatures of silica glass.  The near ubiquity of solid-state emission requires small dust grains that are optically thin, and thus must be replenished on year-to-decade timescales by ongoing collisions.  The disk exhibiting a featureless continuum can only be fit by dust temperatures in excess of 2000\,K, implying highly refractory material comprised of large particles, or non-silicate mineral species.  If confirmed, the glassy silica orbiting two stars could be indicative of high-temperature processes and subsequent rapid cooling, such as occur in high-velocity impacts or vulcanism.  These detections have been enabled by the unprecedented sensitivity of MIRI LRS spectroscopy and highlight the capability and potential for further observations in future cycles.

\end{abstract}


\section{Introduction}

It has been nearly four decades since the confirmation of the first, single white dwarf with infrared excess \citep[G29-38;][]{zuckerman1987}\footnote{The $K$-band photometric excess was first detected by \citet{probst1983} sometime between 1978 and 1980, thus preceding the first historically-recognized detection of an exoplanetary debris disk around Vega with the {\em IRAS} satellite \citep{aumann1984}.}.  Since that time, there have been several key milestones, beginning with the insight that heavy elements in isolated white dwarf atmospheres are accreted from their extant planetary systems \citep[and not the interstellar medium;][]{zuckerman2003}, where G29-38 was the first example with detected photospheric metals and a debris disk \citep{koester1997}.  On the basis of this single object and limited data, it was hypothesized that either a planetesimal collision or catastrophic fragmentation was responsible for the observed circumstellar debris \citep{graham1990,jura2003}.

The launch of the {\em Spitzer Space Telescope} was transformative, revealing a clear correlation between infrared excess from circumstellar dust and metal-enrichment in white dwarf atmospheres \citep{jura2007b,farihi2009a}.  These observations formed the basis for the direct comparison of solar system planetary materials with those observed accreting and diffusing through the outer layers of white dwarfs \citep{zuckerman2007,klein2010}.  During the {\em Spitzer} era, the disks orbiting white dwarfs were found to sometimes exhibit optical emission from co-located metal gas, on top of a dust continuum in the infrared \citep{gaensicke2006,melis2010,brinkworth2012}.  Based on infrared excess detections around roughly three dozen stars,  {\em Spitzer} observations revealed that the dust emission is confined to a region close to the star where the equilibrium temperature is roughly 1000\,K, with no evidence of cooler dust based on MIPS 24\,$\upmu$m observations \citep{farihi2016a}.  

Disk spectroscopy using {\em Spitzer} IRS in low-resolution mode revealed solid-state emission from small grains of silicate minerals toward all eight targets that were successfully acquired during the cryogenic mission \citep{jura2009a}.  These spectroscopic silicate features have distinctive broad shapes and red wings similar to that observed in the solar system zodiacal dust and cometary ejecta \citep{reach2005b,lisse2006}.  These mid-infrared spectroscopic disk studies strengthen the interpretation of polluted white dwarf chemical compositions, and provide in-situ observations of the planetary material that is otherwise unavailable once the debris has been accreted, where it is reduced to atomic components.

There is a consensus that white dwarf debris disks are the result of the catastrophic fragmentation of one or more exoplanetary bodies, but their origins and source regions remain uncertain \citep{bonsor2010,frewen2014,smallwood2018}.  Following the tidal disruption of a planet or planetesimal, the resulting spread and subsequent evolution of particle orbits depend on both the initial size and semimajor axis \citep{malamud2020a,duvvuri2020}, and there are several dynamical processes that might enhance collisions, circularization, and compaction towards the distinct $T\approx1000$\,K emission observed \citep{malamud2021,brouwers2022}.  The subsequent evolution of the closely-orbiting dust disk can be driven by radiation drag on the innermost ring of solids \citep{rafikov2011a,metzger2012}, or via collisions that can enhance gas production and ultimately dominate the mass infall rate \citep{kenyon2017a,kenyon2017b}.

%And while it is possible that a parent body can become chemically-segregated in the process of being disrupted and then later accreted 

The disk material accreted by the white dwarf is an effective mirror for its composition, but it is only one snapshot of this evolution.  And while it has been suggested that a parent body might become chemically segregated in the process of being disrupted and then later accreted \citep{brouwers2023}, there are currently only a few polluted white dwarfs consistent with the accretion of volatile-rich parent bodies \citep{farihi2013c,xu2017,klein2021}.  The largest fraction of polluted white dwarfs are consistent with the accretion of material similar to chondritic meteorites and the bulk Earth \citep{swan2023,doyle2023}, and a moderate subset of accreted bodies that appear differentiated \citep{zuckerman2011,melis2017}.  Observations of the in-situ material are critical to understanding how the atomic bulk compositions were assembled as solids, which provides information on their geochemical and thus planetary origins.  There are currently over 1700 white dwarfs with documented metal pollution \citep{williams2024}, but prior to this letter there were only eight spectroscopic detections of mineralogical dust features \citep{jura2009a}.

In this letter, new observations of 12 polluted white dwarfs are presented using the {\em James Webb Space Telescope (JWST)}, where the low-resolution spectroscopy indicates a detection of the circumstellar disk in all cases.  Section~2 presents the data and analysis, Section~3 discusses the key observational results, and Section~4 concludes with future prospects.

\section{Observations and Analysis}

%%% TABLE %%%
\begin{table*}
\flushleft
\begin{flushleft}
\footnotesize
\caption{Summary of MIRI LRS results\label{tab_main}.}
\begin{tabular}{@{}crlcccccccccc@{}}

\hline
\hline

Star    
&$T_{\rm eff}$
&SpT 
&Distance
&\multicolumn{2}{c}{$\langle {\rm S/N} \rangle$}
&\multicolumn{4}{c}{Disk Emission}
&$T_{\rm dust}$
&$F_{\rm LRS}/F_\star$
&$F_{\rm em}/F_{\rm cont}$\\

&(K)			
&
&(pc)
&[$6\,\upmu$m] 	
&[$10\,\upmu$m] 
&[Optical]	
&[$4\,\upmu$m] 	
&[$6\,\upmu$m]
&[$10\,\upmu$m] 
&(K)
&\multicolumn{2}{c}{[$8-12\,\upmu$m]}\\

\hline

J053753.46$-$475805.1	&23\,000	&DAZ	&92.4       &360	&140	&$-$	&$+$    &$+$	&$+$	&970    &15.1	&0.71\\
J054725.76$-$484722.3	&10\,000	&DZ		&46.5       &170	&29		&$-$	&$-$    &$-$	&$+$	&2040   &0.33	&0.33\\
J064405.23$-$035206.4	&18\,000	&DBZ	&111.9      &400	&150	&$+$	&$+$    &$+$	&$+$	&1220   &26.0	&0.61\\
J070755.11$-$743827.5	&17\,000	&DAZ	&59.0       &260	&35		&$-$	&$+$    &$+$	&$-$	&2040   &0.81	&...\\
J071959.41$+$402122.1	&17\,000	&DBZ	&60.0       &240	&38		&$-$	&$+$    &$+$	&$+$	&850    &1.22	&0.10\\
J072055.81$-$425031.3	&17\,000	&DAZ	&91.6       &100	&14		&$-$	&\nodata&$-$	&$+$	&1200   &0.28	&0.28\\
J080227.74$+$563155.4	&10\,500	&DAZ	&62.7       &390	&290	&$-$	&$+$    &$+$	&$+$	&1180   &28.4	&2.84\\
J084539.18$+$225728.2	&19\,000	&DBZ	&105.5      &330	&82		&$+$	&$+$    &$+$	&$+$	&1230   &11.1	&0.34\\
J084702.28$+$512853.3	&23\,000	&DAZ	&139.0      &250	&150	&$-$	&$+$    &$+$	&$+$	&1230   &25.4	&2.58\\
J085911.29$-$364730.8	&10\,000	&DAZ	&44.1       &250	&120	&$-$	&$+$    &$+$	&$+$	&870    &2.63	&1.62\\
J154144.88$+$645352.9 	&11\,500	&DAZ	&55.8       &650	&230	&$-$	&$+$    &$+$	&$+$	&980    &15.6	&0.04\\
J161316.63$+$552125.9	&12\,000	&DAZ	&65.9       &300	&340	&$-$	&$+$    &$+$	&$+$	&970    &36.3	&6.92\\

\hline

\normalsize
\end{tabular}
\end{flushleft} 

{\bf Note}.  $T_{\rm eff}$ is that adopted in the process of fitting atmospheric models to short-wavelength photometry (Section~2 and Figure~\ref{fig_app}).  From left to right, the columns for disk emission refer to those apparent in the optical via gas lines, as thermal continuum in warm {\em Spitzer} IRAC or equivalent {\em WISE} photometry, at $6\,\upmu$m in flux-calibrated LRS spectra, or via solid-state emission around $10\,\upmu$m in LRS.  $T_{\rm dust}$ is the best fit value to the thermal continuum of the excess photometric flux.  The last two columns are the summed fluxes of each component over $8-12\,\upmu$m.

\end{table*}

All data in this letter were taken as part of Cycle 2 Survey program 3690 (PI: Farihi).  Survey targets were selected to be known, single metal-polluted white dwarfs via published work or ongoing efforts, with $G<17.0$\,mag and generally restricted to $T_{\rm eff}\la25\,000$\,K, where the effects of radiative levitation cannot account for their photospheric heavy elements \citep{schatzman1945}.  The survey was agnostic to all other stellar parameters (e.g.\ distance, primary atmospheric composition, metal abundance, diffusion timescale), and to whether or not the systems are known to possess infrared excesses.

Altogether, 36 metal-polluted white dwarfs were observed with the Mid-Infrared Instrument \citep[MIRI;][]{rieke2015} aboard {\em JWST}, using the Low-Resolution Spectrometer (LRS) in slit mode.  This MIRI module provides spectroscopy through a 0\farcs51$\times$4\farcs7 slit, with resolving power $R\approx100$ over the wavelength range $5.0-12.5\,\upmu$m.  Acquisition was carried out using the target itself, with the F560W broad-band imaging filter, ten groups and fast readout in a single integration of 28\,s, followed by a verification image to ensure the target was properly centered at the slit, using identical instrument settings.  Spectroscopy was executed by nodding along the slit, with two dithers consisting of three integrations each, where a single integration consisted of 70 groups and a fast readout.  The total on-source integration time for this spectroscopic setup was 1177\,s.  Owing to this fixed time spent on all targets in the survey, the resulting S/N varies based on source brightness.

The raw spectral data (\texttt{uncal} files) were processed and calibrated with the {\em JWST} Science Calibration Pipeline \citep{bushouse2023} version 1.14, with the reference files in CRDS context \texttt{jwst\_1231.pmap}.  The calibration factors (PHOTMJSR, conversions from DN\,pixel$^{-1}$ to MJy\,sr$^{-1}$) are 14.71 and 0.455 for the LRS slit and target acquisition imaging modes, respectively. Standard reduction steps were used to produce the final extracted, 1-D spectra including two-nod background subtraction and spectral extraction using a fixed width of 8\,pixels. Similarly, the pipeline produced source catalog was adopted for the target acquisition image, and the F560W photometry was automatically extracted.  However, the F560W filter fully overlaps the LRS wavelength range, and the acquisition images are not of photometric quality as they contain cosmic rays and bad pixels.  Nevertheless, all acquisition images were scrutinized and found to contain a single point source and no indication of background or neighboring sources at the position of the target. The spectroscopic errors are the result of the standard pipeline extractions, which include measurement errors added in quadrature with the instrument calibration error.

Table~\ref{tab_main} gives a brief summary of the targets with infrared excess detected by MIRI in this work, and provides estimated signal-to-noise (S/N) ratios over a $1\,\upmu$m interval centered at both 6 and $10\,\upmu$m.  The table also indicates if the target has disk emission detected in the optical (via gas), thermal continuum emission in the 4 and $6\,\upmu$m regions, or a silicate emission feature at $10\,\upmu$m.  Each target acquisition image shows a single source at the position of the white dwarf within 0\farcs5, and the pipeline extracted photometry agrees well with each LRS spectrum.

For each target with data, short-wavelength photometry was sourced from catalogs provided by facilities such as {\em GALEX}, Pan-STARRS, SkyMapper, 2MASS, and {\em WISE} \citep{martin2005,skrutskie2006,keller2007,wright2010,tonry2012}.  The optical and near-infrared data below 2\,$\upmu$m were used to fit an appropriate, pure hydrogen or helium white dwarf atmosphere model \citep{koester2010}.  Ultraviolet fluxes were visualized, but typically ignored in the fitting process owing to both extinction and the unknown extent of metal absorption in this wavelength region.  Similarly, the infrared fluxes from {\em WISE} were used to evaluate the fitting process, but not as photometric anchors for fitting stellar atmospheres, as these data often exhibit flux contamination from neighboring or background sources, and low signal-to-noise for white dwarfs \citep{farihi2016a,dennihy2020}.  The model atmospheres fitted to the photometric data are shown in Figure~\ref{fig_app}.

The {\em JWST} MIRI spectra are flux-calibrated to around 2\% or better at $6-7\,\upmu$m\footnote{\url{https://jwst-docs.stsci.edu/jwst-calibration-status}}, and thus the LRS data can be directly compared with the extrapolated white dwarf photosphere at the relevant wavelengths.  Of the 36 metal-enriched, single stars observed in the Cycle 2 Survey, there are 12 unambiguous detections of infrared excess above the stellar continua in their LRS spectra, where the threshold for a confident excess is around 8\%, comprising the instrument calibration error and another 2\% uncertainty for the modeled stellar flux at these wavelengths.  The weakest excesses detected are approximately 30\% larger than the predicted stellar photosphere at $8-12\,\upmu$m (Table~\ref{tab_main}).  

For the stars with detected infrared excesses, the dust temperatures were estimated using blackbody fits to available infrared photometry in the $2-5\,\upmu$m range, together with segments of the LRS spectra unaffected by solid-state emission (typically within $5-8\,\upmu$m).  These dust continuum temperatures are listed as $T_{\rm dust}$ in Table~\ref{tab_main}, and have characteristic uncertainties of $\pm100$\,K, but up to $\pm200$\,K in the hottest cases.  The remaining 24 targets have LRS spectra that are consistent with the stellar photosphere at all wavelengths up to the $12\,\upmu$m, and will be presented in a future paper.

\section{Disk Detections}

\begin{figure*}
\includegraphics[width=0.33\textwidth]{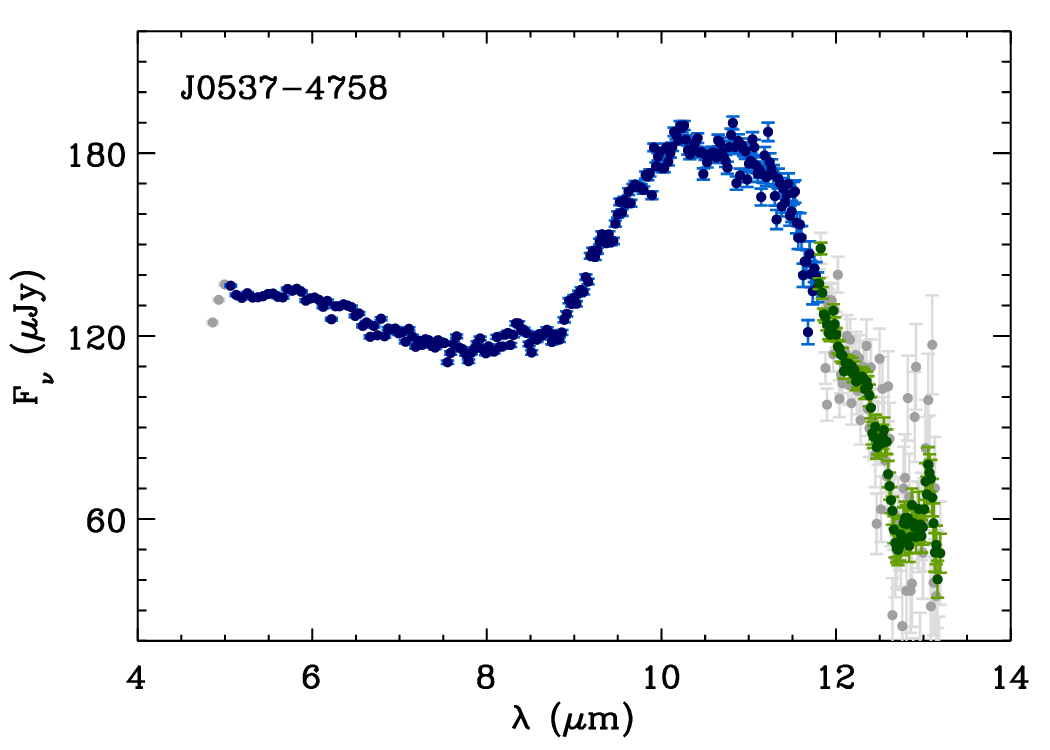}
\includegraphics[width=0.33\textwidth]{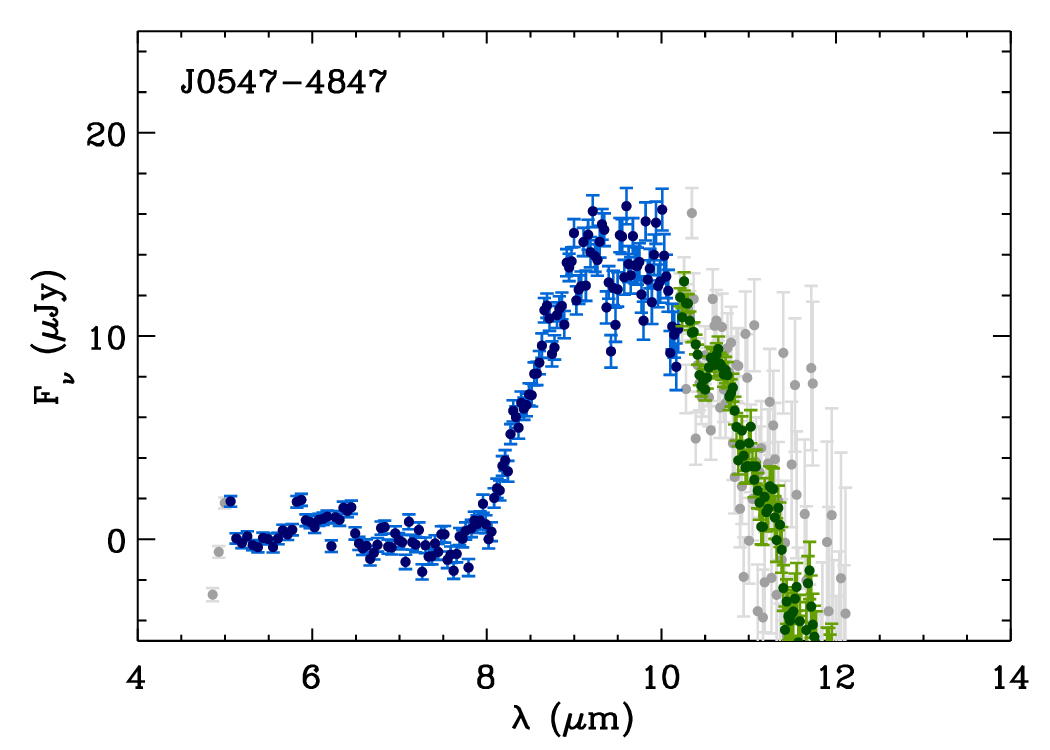}
\includegraphics[width=0.33\textwidth]{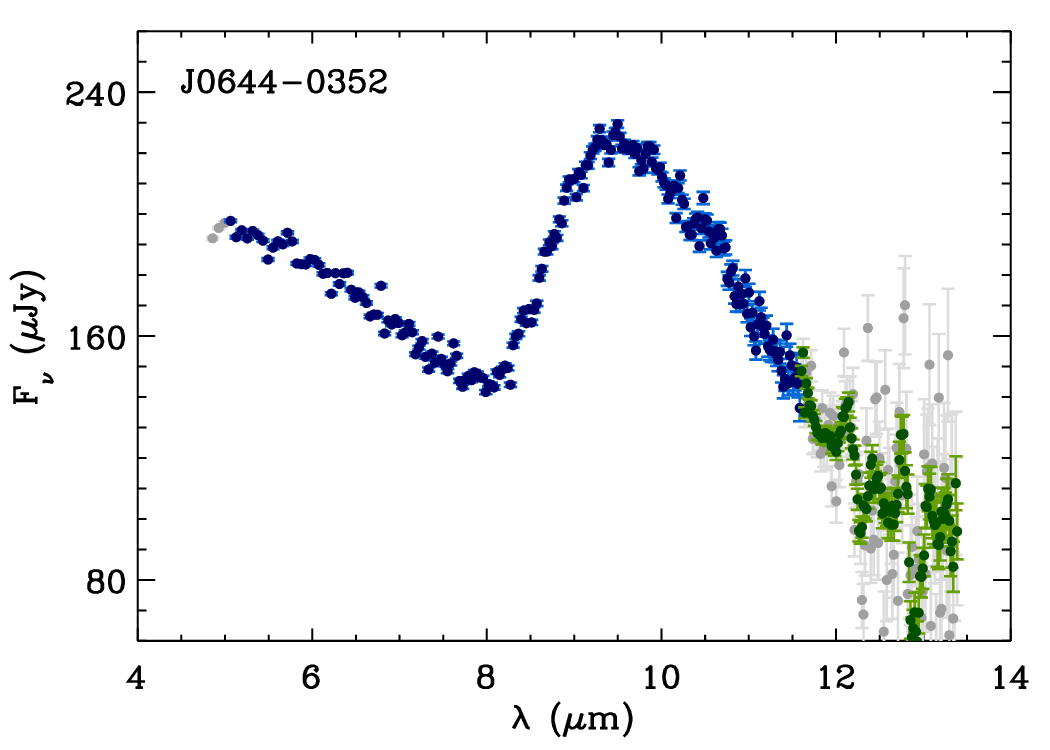}
\includegraphics[width=0.33\textwidth]{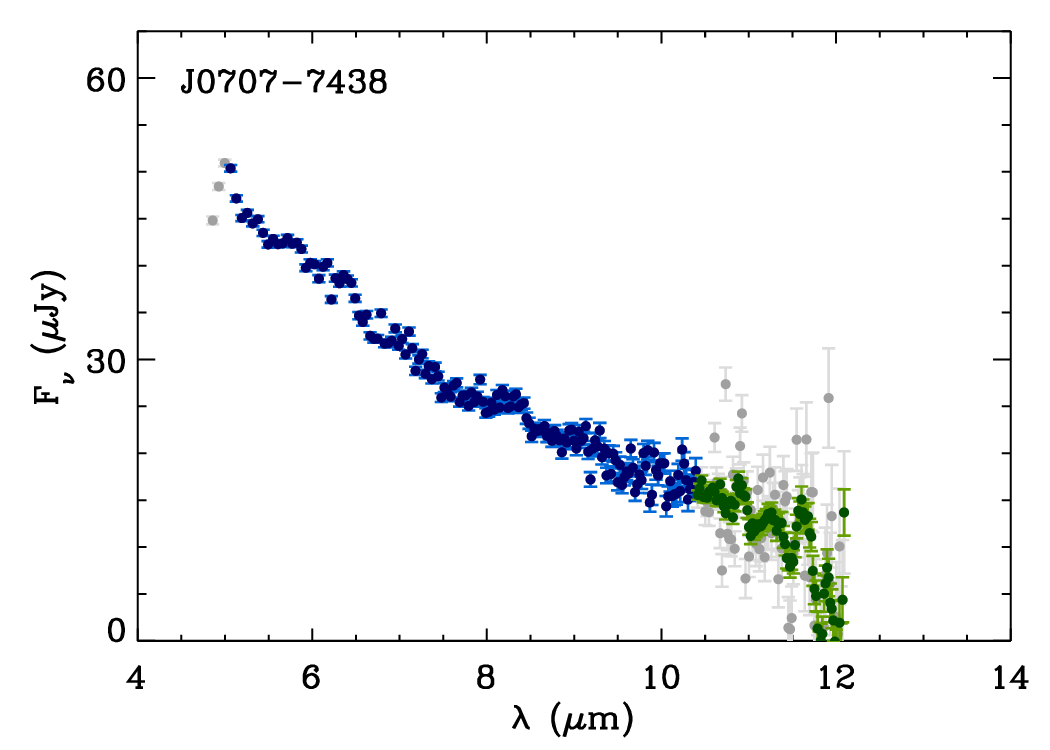}
\includegraphics[width=0.33\textwidth]{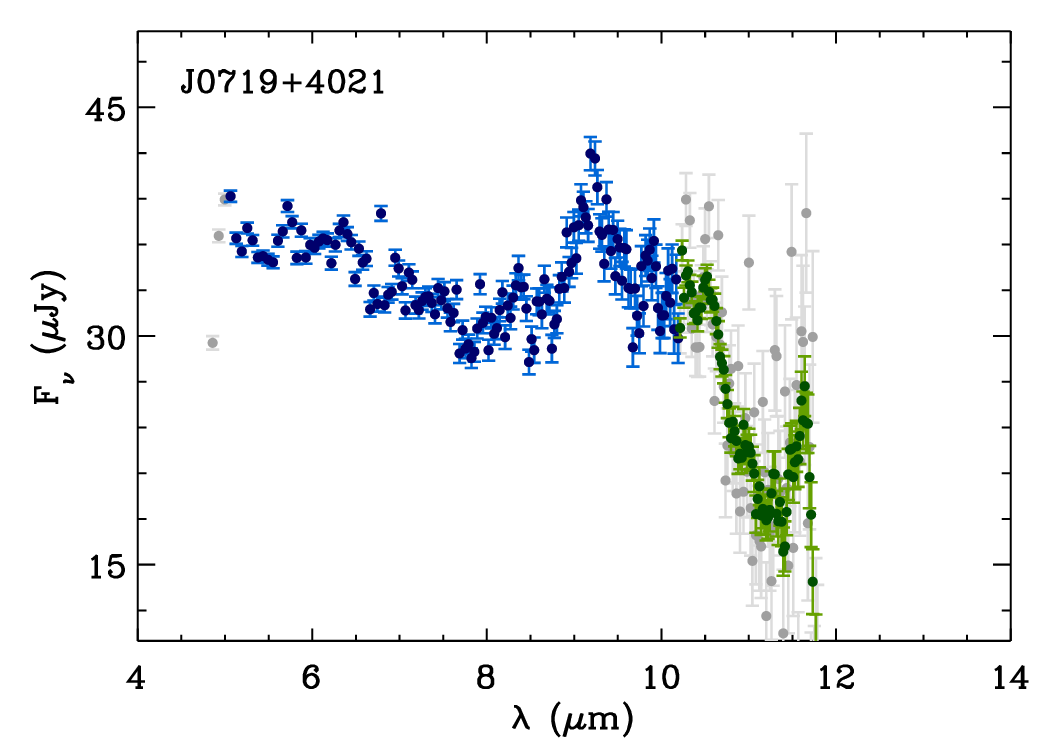}
\includegraphics[width=0.33\textwidth]{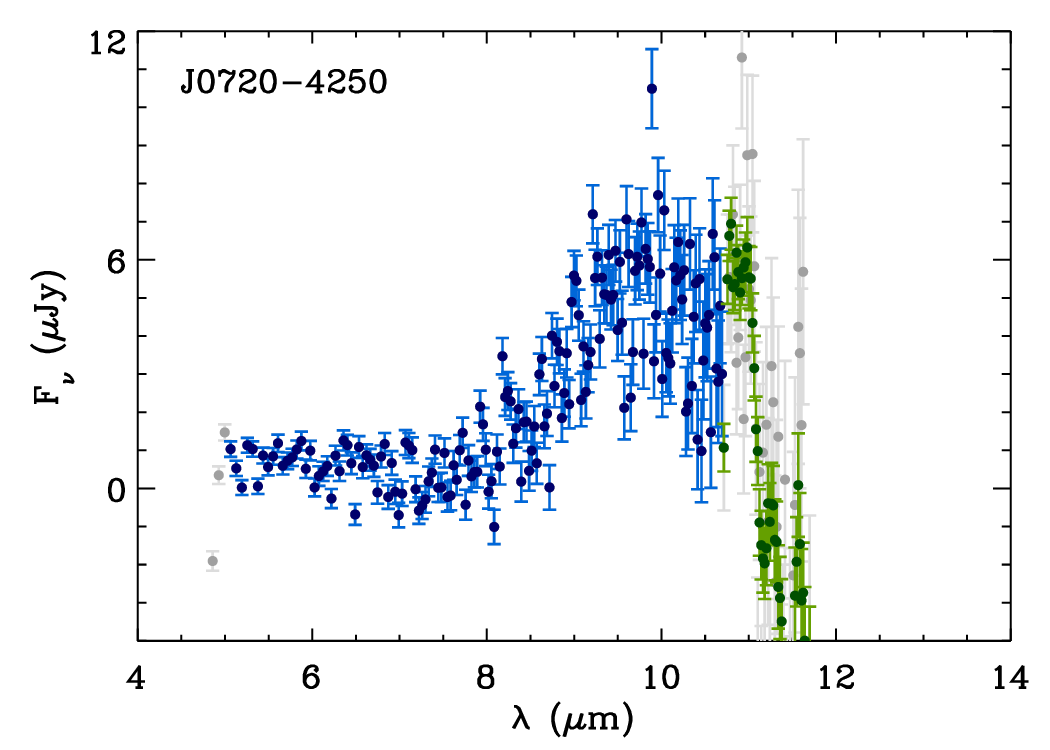}
\includegraphics[width=0.33\textwidth]{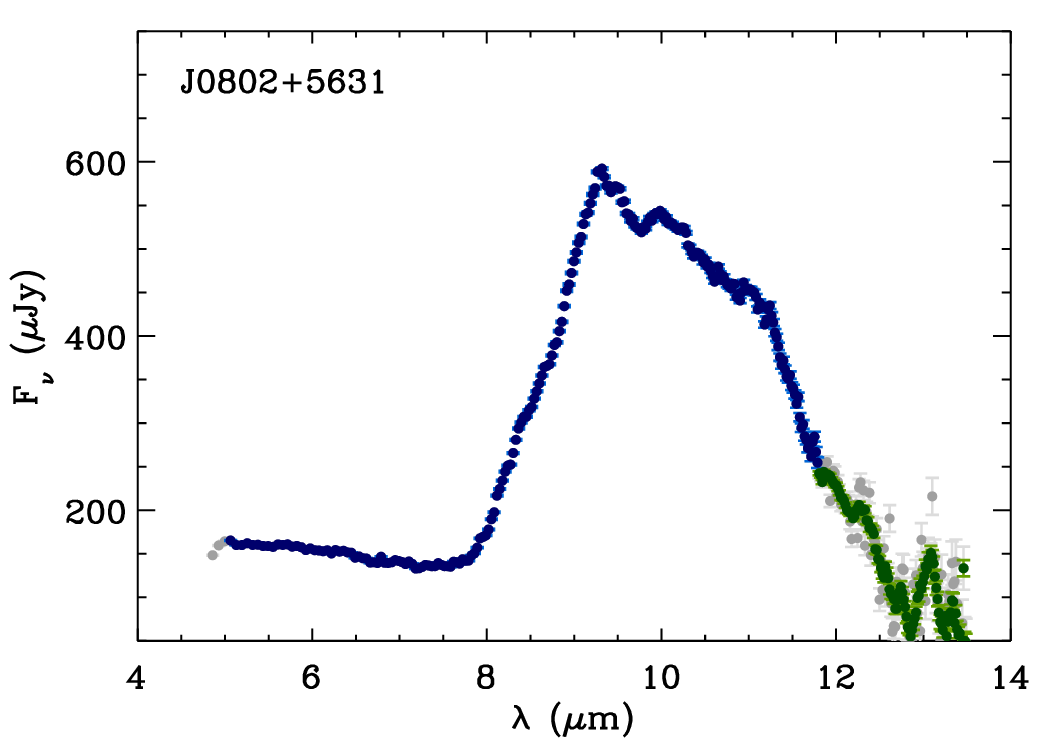}
\includegraphics[width=0.33\textwidth]{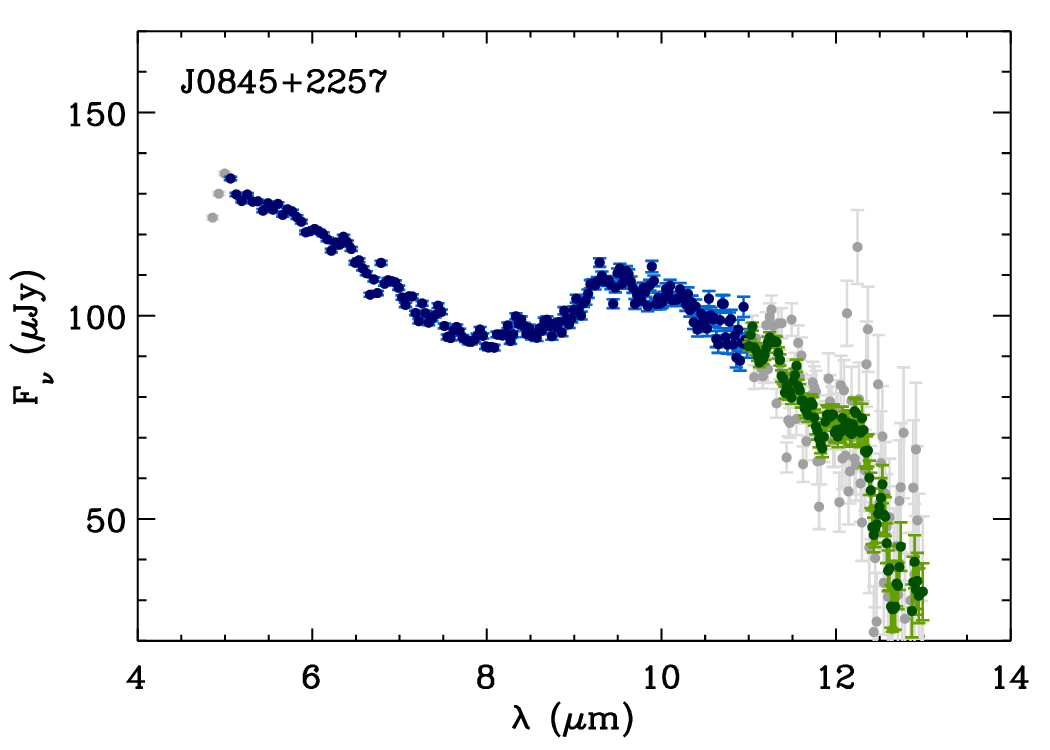}
\includegraphics[width=0.33\textwidth]{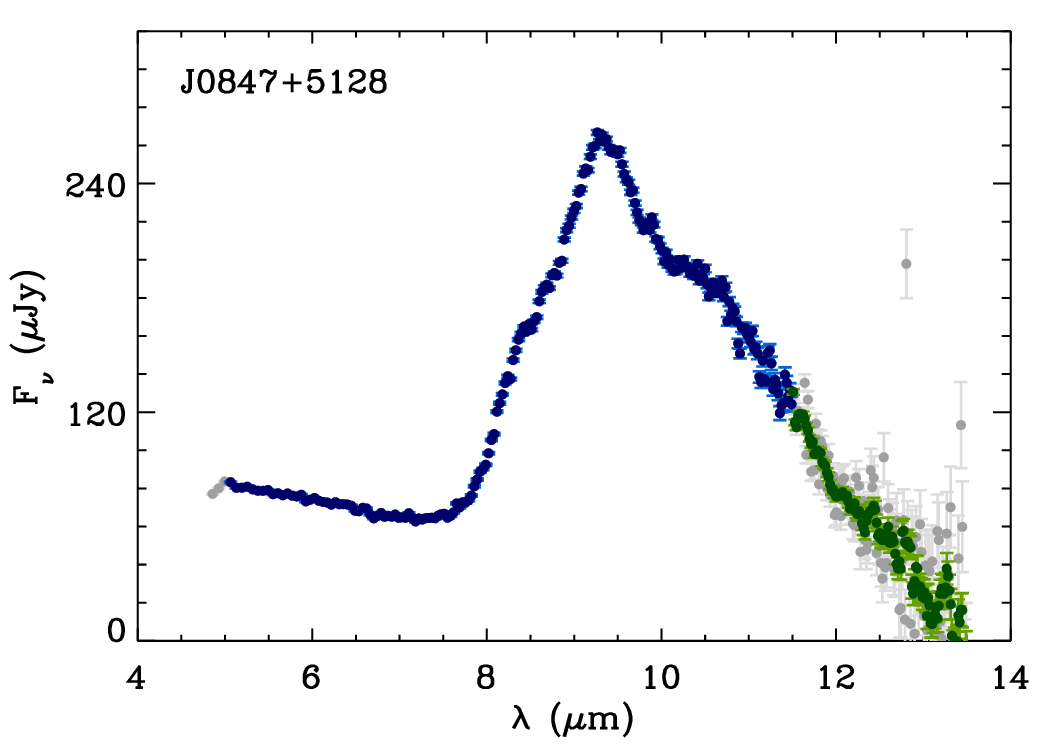}
\includegraphics[width=0.33\textwidth]{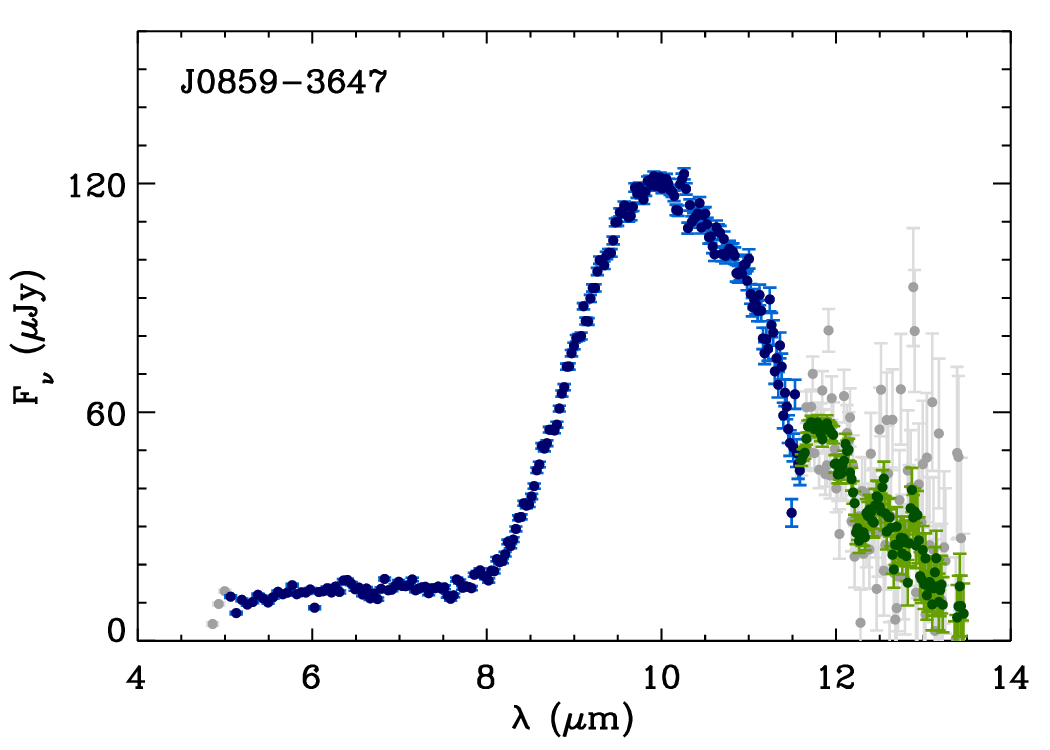}
\includegraphics[width=0.33\textwidth]{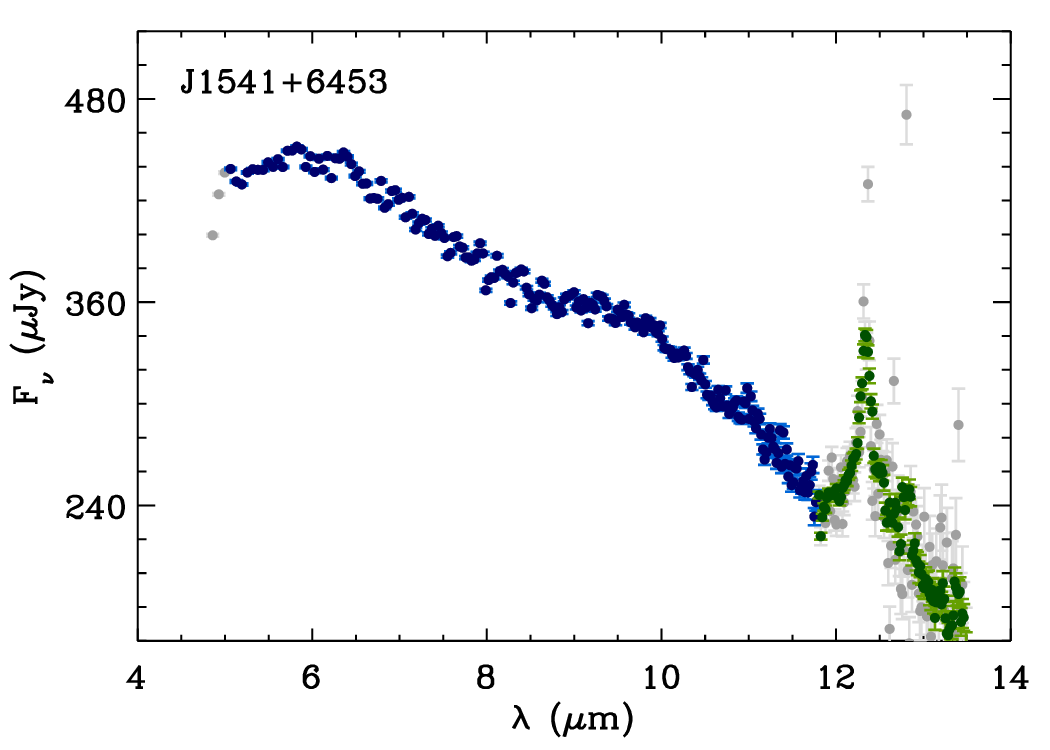}
\includegraphics[width=0.33\textwidth]{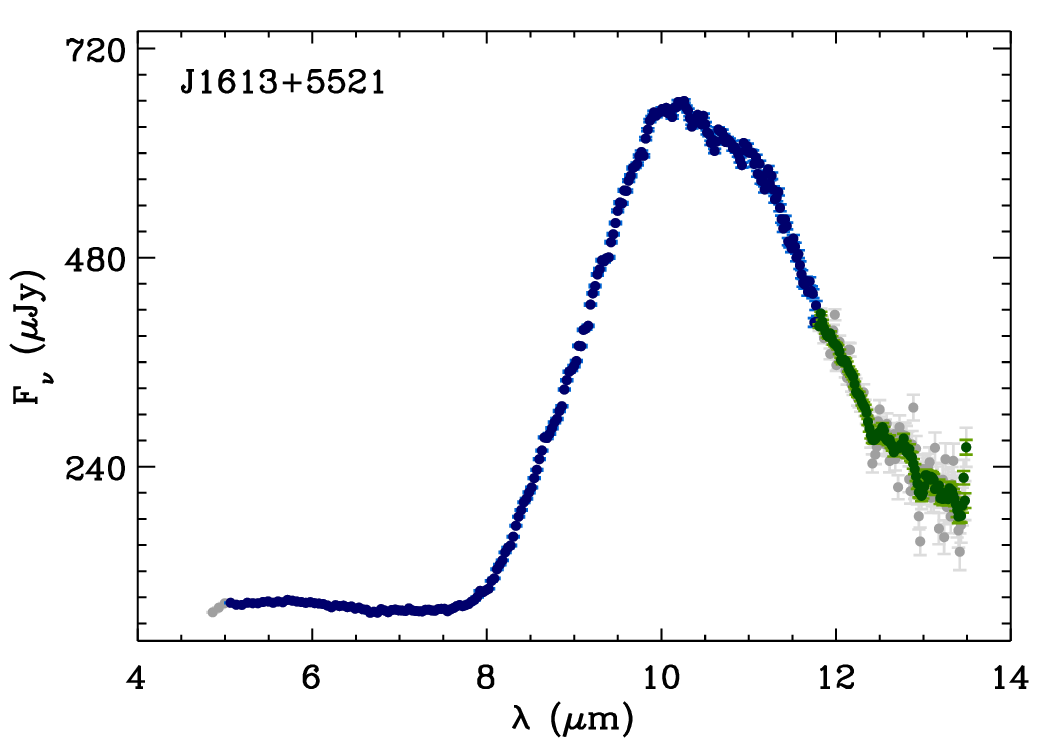}
\vskip 0pt
\caption{Gallery of {\em JWST} MIRI LRS spectra for 12 targets observed in Cycle 2 Survey program 3690.  These are the dust fluxes with the stellar atmosphere subtracted, plotted as data points with error bars.  The processed and extracted data are shown in grey, and these are overplotted in blue for regions where features are not obscured by the scatter in flux, and in green using 7-pixel boxcar smoothing for the noisiest regions.  Highlights include 1) the subtle detections of disks via weak silicate features in J0547$-$4847 and J0720$-$4250, 2) the lack of any silicate emission in J0707$-$7438, 3) the blue-peaked solid-state features in J0719$+$4021, J0802$+$5631, and J0847$+$5128, which are consistent with glassy silica such as found in tektites and obsidian, and 4) the strongest silicate feature observed to date in any white dwarf; J1613$+$5521 is an order of magnitude stronger (see Figure~\ref{fig_vs}) than the feature observed towards G29-38, the dusty white dwarf prototype \citep{reach2009}.  The narrow feature near $12\,\upmu$m in J1541$+$6453 is almost certainly due to a cosmic ray, as it is seen in only one of the two LRS nod positions.
\label{fig_main}}
\end{figure*}

\begin{figure}
\includegraphics[width=\columnwidth]{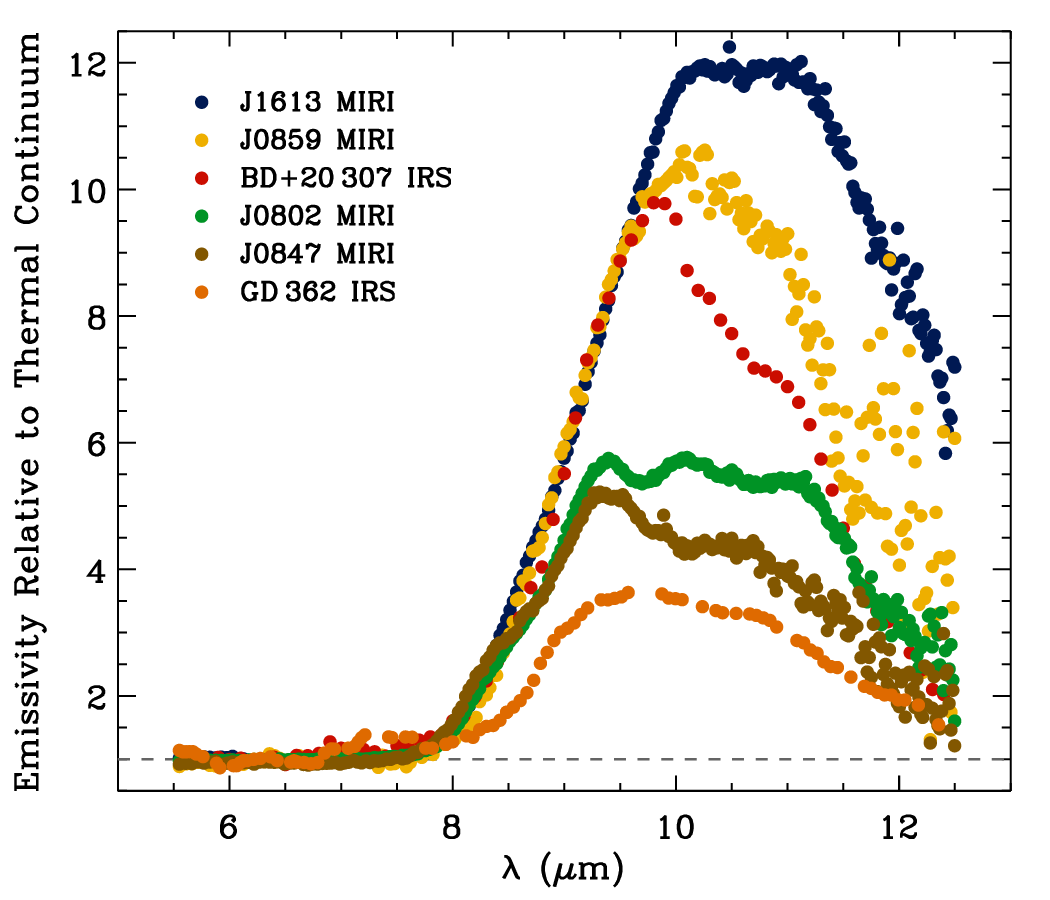}
\caption{The four strongest silicate features observed with {\em JWST} MIRI, as compared with {\em Spitzer} IRS spectra for the previous record holder for a white dwarf debris disk \citep[GD\,362, now the weakest in the plot;][]{jura2009a}, and the brightest silicate emission seen toward a dusty main-sequence star \citep[BD+20\,307;][]{song2005,weinberger2011}.  The data are plotted as the emissivity relative to the fitted thermal continuum at each wavelength (see Section~2 for details), where the dashed grey line represents a blackbody with an emissivity of unity everywhere.  The silicate emission detected toward GD\,362 in 2006 \citep{jura2007a} is roughly twice as strong as that observed around G29-38 \citep{reach2005b}, and thus the silicate features seen in the disks of J0859$-$3647 and and J1613$+$5521 are six to seven times stronger than for the dusty white dwarf prototype, and are among the strongest observed in any debris disk orbiting any main-sequence or white dwarf star. 
\label{fig_vs}}
\end{figure}

For each of the 12 stars with infrared excess in their LRS spectra, the fitted white dwarf photospheric continuum (Section~2) was subtracted from the total infrared fluxes and the resulting dust-only spectra are plotted in Figure~\ref{fig_main}.  While the group of disk spectra have common traits, there is substantial diversity and several novelties.  

Ten systems have easily seen solid-state emission features in the $9-11\,\upmu$m region that arise from small grains of silicate mineral species; an eleventh object (J1541$+$6453) also has a weak feature that is more readily visualized after removal of the strong thermal continuum.  The emission features exhibit a wide range of intrinsic brightnesses over the $8-12\,\upmu$m interval, where they span over two orders of magnitude as flux ratios relative to the stellar and dust continua (see Table~1).  Solid-state emission arises only from optically thin dust regions, but alone cannot distinguish between disks dominated by low or high optical depths over their surface \citep{reach2009,jura2009a}.  However, the requirement of small grain sizes does imply these have been produced recently.  Poynting-Robertson drag clears micron-sized dust within 1\,R$_\odot$ in less than $1-10$\,yr for typical white dwarf luminosities \citep{rafikov2011a}, and therefore this dust must be the product of ongoing collisions \citep{kenyon2017a,malamud2021,swan2021}.

The disk orbiting J0707$-$7438 has no indication of small silicate grains, and by itself the naked thermal continuum might indicate an absence of these common minerals (e.g.\ dominated by metallic iron and alloys), or alternatively, silicate grains too large to result in detectable emission.  While amorphous carbon is also possible, it is well-established that the accreted materials in polluted white dwarfs are carbon poor \citep{wilsond2016}.  For this star, the continuum excess is well established both by the MIRI flux-calibrated data as well as {\em WISE} photometry at 3.4 and 4.6\,$\upmu$m; the two datasets independently support an excess flux consistent with a single temperature blackbody across their individual wavelengths.  Furthermore, the white dwarf is relatively isolated in the {\em WISE} image fields and should have minimal or no contamination, and only a single point source is seen in the MIRI target acquisition images.  

Fitting a blackbody to the excess flux yields a dust temperature of 2040\,K, indicating refractory material in close proximity to the white dwarf, where it is possible for metallic iron to attain such high temperatures in a hydrogen-poor environment \citep{rafikov2012,steckloff2021}.  This may be the hottest dust temperature inferred for any main-sequence or white dwarf star to date.  While a relatively featureless brown dwarf companion \citep{miles2023} is difficult to rule out definitively without reliable $K$-band photometry or spectroscopy, this possibility is unlikely for a few reasons: i) brown dwarfs are rarely companions to main-sequence stars and white dwarfs \citep{farihi2005b,grether2006}, ii) polluted white dwarfs are dominated by single stars \citep{noor2024}, iii) the F560W acquisition images rule out an additional source beyond 4\,AU in projected separation, and iv) the {\em Gaia} RUWE is consistent with a single object.

There are three examples (J0719$+$4021, J0802$+$5631, J0847$+$5128) where glassy or amorphous silica dust is strongly suggested by the sharp peak in their emission features near 9\,$\upmu$m, analogous to the stronger feature in the debris disk orbiting the main-sequence star HD\,172555 \citep{lisse2009}.  Two plausible candidates for this material are extrasolar versions of tektites and obsidian, which are both terrestrial materials that are formed near the surface of the Earth at high temperatures with subsequent rapid cooling.  Whereas tektites are high-velocity impact ejecta with almost no volatile or crystal content, obsidian and similar volcanic glasses have modest water content and some small crystals inclusions.  

However, two other possibilities for such glassy silica are discussed in the literature \citep{lisse2020}.  One is rapid vaporization and re-condensation of SiO gas in high-energy stellar flares.  While all white dwarfs are forged at extremely hot temperatures, the volume that is heated above 1000\,K remains well within 1\,AU, and there may be no opportunity for re-condensation of any vaporized silicates.  Another possibility is space weathering, which preferentially affects Fe-silicates but may require abundant water ice to create the necessary SiO gas that can potentially recondense as glass.  All the aforementioned scenarios require detailed mineralogical modeling and are beyond the scope of this letter, but it is noteworthy that high-velocity impacts may be relatively common in white dwarf debris disks, given that the circular Keplerian speed is 330\,km\,s$^{-1}$ at 1\,$R_\odot$ around a 0.6\,$M_\odot$ host star.

In terms of fractional infrared luminosity, this survey has uncovered the two faintest debris disk detections by an order of magnitude \citep[J0547$-$4847 and J0720$-$4250;][]{farihi2010b,rocchetto2015}.  Furthermore, J0547$-$4847 is the first white dwarf with spectral type DZ (metal lines only) to have an unambiguous infrared excess, which is 30\% above the stellar photosphere.  This star is likely to have a helium-dominated atmosphere with heavy element sinking timescales on the order of Myr, and is an important detection as few white dwarfs with $T_{\rm eff}\la10\,000$\,K are known to have infrared excess detected by any means prior to {\em JWST} \citep{bergfors2014,melis2020}.  

Perhaps most intriguing is that there is a lack of detectable excess from a thermal dust continuum in these two faint disks, where the presence of dust is only revealed via their subtle silicate features.  In both cases, the $6-7\,\upmu$m LRS fluxes are within 2\% (on average) of the stellar model predictions, and thus below the estimated 8\% threshold for a confident excess.  These are the first debris disks detected without significant $T\approx1000$\,K continuum emission that has been the hallmark of dusty white dwarfs in the eras of {\em Spitzer} and {\em WISE} \citep{farihi2009a,lai2021}, and thus only detectable via infrared spectroscopy beyond $8\,\upmu$m.  It is likely that the most subtle disk detections represent at least 15\% of previously-undetected disks (considering only those stars observed with LRS that had no prior-known infrared excess).

Among the dust spectra are two objects with possibly the strongest silicate features observed to date around any mature star (J0859$-$3647 and J1613$+$5521).  In Figure~\ref{fig_vs}, the four brightest emission features detected in this survey are compared with the previously known strongest features in both white dwarf and main-sequence debris disks, and plotted as emissivities normalized to the thermal continua at each wavelength.  For each of these four stars, the photosphere was modeled based on published stellar parameters and fitted to short wavelength photometry, and subsequently subtracted from each MIRI spectrum and available infrared photometry from {\em Spitzer} and {\em WISE}.  The featureless continuum of the resulting infrared excess flux was modeled using a single blackbody for the MIRI targets, but in the case of GD\,362 and BD+20\,307 where photometry or spectroscopy is available at longer wavelengths, an optimal fit to the continuum was achieved with two blackbodies.  The plotted emissivities are the ratio of the dust spectra to the fitted thermal continua, which are unity at all wavelengths by definition for blackbody spectra.

It should be noted that the four stars with newly discovered strong silicate features currently lack longer wavelength photometry, which limits the ability to model the thermal dust continuum (e.g.\ as two blackbodies), and establish more accurate emissivities.  Thus, the emissivities in Figure~\ref{fig_vs} are indicative but perhaps not definitive.  Naively comparing J1613$+$5521 with G29-38 and assuming all else equal, the dust mass of optically thin particles should be on the order of $3\times10^{19}$\,g \citep{ballering2022}.  This spectacular feature is sufficiently bright to be one of few detected in {\em WISE} 12\,$\upmu$m imaging, and indicates ongoing collisional activity that is producing sub-micron sized particles that would otherwise be removed within a few years.

\section{Summary and Future Work}

This survey program has uncovered 12 of 36 polluted white dwarfs with excess, and this detection rate is at least $3\times$ better than that achieved with {\em Spitzer} for a similarly biased sample of stars with and without known photometric excess \citep[e.g.][]{mullally2007,debes2007,jura2007b,farihi2008a}.  The spectra are diverse, representing both the weakest and strongest dust emission yet observed toward any white dwarf, and with novel morphologies and thus mineral compositions not previously hinted at with {\em Spitzer} IRS.  

Future work requires modeling of the solid-state features to determine the particular mineral species compositions of the silicate dust, where this may be compared to what is observed in the atmosphere.  However, caution is warranted as the silicate features are likely to contain only $\sim10^{18\pm1}$\,g of dust \citep{reach2009,ballering2022}, whereas it is well established that many white dwarfs with sufficiently long sinking timescales have accreted up to $\sim10^{23}$\,g or more in total heavy elements  \citep{hollands2018a,swan2023}.  Thus, the mineral content of dust observed in-situ can easily be dwarfed by the bulk of the debris accreted onto the stellar surface over time.  Analysis of the full survey results will be presented in a forthcoming paper, including LRS spectra for the 24 metal-polluted targets with no indications of excess dust emission.  Both modeling and additional observations are required to better understand the link between the debris disk and polluted atmosphere.

\section*{acknowledgments}
The authors thank J.~K.~Steckloff for clarification on models of high temperature dust.  This work is based on observations made with the NASA / ESA / CSA {\em James Webb Space Telescope}.  The authors acknowledge support from NASA grant JWST-GO-03690, and KYLS acknowledges partial support from NASA XRP grant 80NSSC22K0234.  The data were obtained from the Mikulski Archive for Space Telescopes at the STScI, which is operated by the Association of Universities for Research in Astronomy, under NASA contract NAS 5-03127 for {\em JWST}. These observations are associated with program 3690, and can be accessed via~\dataset[10.17909/xpr9-e989]{http://dx.doi.org/10.17909/xpr9-e989} DOI: 10.17909/xpr9-e989.
%\end{acknowledgments}


\vspace{5mm}

\facilities{JWST(MIRI)}


%\software{Any?  Thank you IDL!}

\bibliography{../../../references}{}

\begin{thebibliography}{}
\expandafter\ifx\csname natexlab\endcsname\relax\def\natexlab#1{#1}\fi
\providecommand{\url}[1]{\href{#1}{#1}}
\providecommand{\dodoi}[1]{doi:~\href{http://doi.org/#1}{\nolinkurl{#1}}}
\providecommand{\doeprint}[1]{\href{http://ascl.net/#1}{\nolinkurl{http://ascl.net/#1}}}
\providecommand{\doarXiv}[1]{\href{https://arxiv.org/abs/#1}{\nolinkurl{https://arxiv.org/abs/#1}}}

\bibitem[{{Aumann} {et~al.}(1984){Aumann}, {Gillett}, {Beichman}, {de Jong},
  {Houck}, {Low}, {Neugebauer}, {Walker}, \& {Wesselius}}]{aumann1984}
{Aumann}, H.~H., {Gillett}, F.~C., {Beichman}, C.~A., {et~al.} 1984, \apjl,
  278, L23, \dodoi{10.1086/184214}

\bibitem[{{Ballering} {et~al.}(2022){Ballering}, {Levens}, {Su}, \&
  {Cleeves}}]{ballering2022}
{Ballering}, N.~P., {Levens}, C.~I., {Su}, K. Y.~L., \& {Cleeves}, L.~I. 2022,
  \apj, 939, 108, \dodoi{10.3847/1538-4357/ac9a4a}

\bibitem[{{Bergfors} {et~al.}(2014){Bergfors}, {Farihi}, {Dufour}, \&
  {Rocchetto}}]{bergfors2014}
{Bergfors}, C., {Farihi}, J., {Dufour}, P., \& {Rocchetto}, M. 2014, \mnras,
  444, 2147, \dodoi{10.1093/mnras/stu1565}

\bibitem[{{Bonsor} \& {Wyatt}(2010)}]{bonsor2010}
{Bonsor}, A., \& {Wyatt}, M. 2010, \mnras, 409, 1631,
  \dodoi{10.1111/j.1365-2966.2010.17412.x}

\bibitem[{{Brinkworth} {et~al.}(2012){Brinkworth}, {G{\"a}nsicke}, {Girven},
  {Hoard}, {Marsh}, {Parsons}, \& {Koester}}]{brinkworth2012}
{Brinkworth}, C.~S., {G{\"a}nsicke}, B.~T., {Girven}, J.~M., {et~al.} 2012,
  \apj, 750, 86, \dodoi{10.1088/0004-637X/750/1/86}

\bibitem[{{Brouwers} {et~al.}(2022){Brouwers}, {Bonsor}, \&
  {Malamud}}]{brouwers2022}
{Brouwers}, M.~G., {Bonsor}, A., \& {Malamud}, U. 2022, \mnras, 509, 2404,
  \dodoi{10.1093/mnras/stab3009}

\bibitem[{{Brouwers} {et~al.}(2023){Brouwers}, {Bonsor}, \&
  {Malamud}}]{brouwers2023}
---. 2023, \mnras, 519, 2646, \dodoi{10.1093/mnras/stac3316}

\bibitem[{{Bushouse} {et~al.}(2023){Bushouse}, {Eisenhamer}, {Dencheva},
  {Davies}, {Greenfield}, {Morrison}, {Hodge}, {Simon}, {Grumm}, {Droettboom},
  {Slavich}, {Sosey}, {Pauly}, {Miller}, {Jedrzejewski}, {Hack}, {Davis},
  {Crawford}, {Law}, {Gordon}, {Regan}, {Cara}, {MacDonald}, {Bradley},
  {Shanahan}, {Jamieson}, {Teodoro}, {Williams}, \&
  {Pena-Guerrero}}]{bushouse2023}
{Bushouse}, H., {Eisenhamer}, J., {Dencheva}, N., {et~al.} 2023, {JWST
  Calibration Pipeline}, 1.12.5,  Zenodo, \dodoi{10.5281/zenodo.10022973}

\bibitem[{{Debes} {et~al.}(2007){Debes}, {Sigurdsson}, \& {Hansen}}]{debes2007}
{Debes}, J.~H., {Sigurdsson}, S., \& {Hansen}, B. 2007, \aj, 134, 1662,
  \dodoi{10.1086/521394}

\bibitem[{{Dennihy} {et~al.}(2020){Dennihy}, {Farihi}, \&
  {Fusillo}}]{dennihy2020}
{Dennihy}, E., {Farihi}, J., \& {Fusillo}, Nicola Pietro Gentile~{Debes}, J.~H.
  2020, \apj, 891, 97, \dodoi{10.3847/1538-4357/ab7249}

\bibitem[{{Doyle} {et~al.}(2023){Doyle}, {Klein}, {Dufour}, {Melis},
  {Zuckerman}, {Xu}, {Weinberger}, {Trierweiler}, {Monson}, {Jura}, \&
  {Young}}]{doyle2023}
{Doyle}, A.~E., {Klein}, B.~L., {Dufour}, P., {et~al.} 2023, \apj, 950, 93,
  \dodoi{10.3847/1538-4357/acbd44}

\bibitem[{{Duvvuri} {et~al.}(2020){Duvvuri}, {Redfield}, \&
  {Veras}}]{duvvuri2020}
{Duvvuri}, G.~M., {Redfield}, S., \& {Veras}, D. 2020, \apj, 893, 166,
  \dodoi{10.3847/1538-4357/ab7fa0}

\bibitem[{{Farihi}(2016)}]{farihi2016a}
{Farihi}, J. 2016, \nar, 71, 9, \dodoi{10.1016/j.newar.2016.03.001}

\bibitem[{{Farihi} {et~al.}(2010){Farihi}, {Barstow}, {Redfield}, \&
  {Dufour}}]{farihi2010b}
{Farihi}, J., {Barstow}, M.~A., {Redfield}, S., \& {Dufour}, P.~{Hambly}, N.~C.
  2010, \mnras, 404, 2123, \dodoi{10.1111/j.1365-2966.2010.16426.x}

\bibitem[{{Farihi} {et~al.}(2005){Farihi}, {Becklin}, \&
  {Zuckerman}}]{farihi2005b}
{Farihi}, J., {Becklin}, E.~E., \& {Zuckerman}, B. 2005, \apjs, 161, 394,
  \dodoi{10.1086/444362}

\bibitem[{{Farihi} {et~al.}(2008){Farihi}, {Burleigh}, \&
  {Hoard}}]{farihi2008a}
{Farihi}, J., {Burleigh}, M.~R., \& {Hoard}, D.~W. 2008, \apj, 674, 421,
  \dodoi{10.1086/524933}

\bibitem[{{Farihi} {et~al.}(2013){Farihi}, {G{\"a}nsicke}, \&
  {Koester}}]{farihi2013c}
{Farihi}, J., {G{\"a}nsicke}, B.~T., \& {Koester}, D. 2013, Science, 342, 218,
  \dodoi{10.1126/science.1239447}

\bibitem[{{Farihi} {et~al.}(2009){Farihi}, {Jura}, \&
  {Zuckerman}}]{farihi2009a}
{Farihi}, J., {Jura}, M., \& {Zuckerman}, B. 2009, \apj, 694, 805,
  \dodoi{10.1088/0004-637X/694/2/805}

\bibitem[{{Frewen} \& {Hansen}(2014)}]{frewen2014}
{Frewen}, S.~F.~N., \& {Hansen}, B.~M.~S. 2014, \mnras, 439, 2442,
  \dodoi{10.1093/mnras/stu097}

\bibitem[{{G{\"a}nsicke} {et~al.}(2006){G{\"a}nsicke}, {Marsh}, {Southworth},
  \& {Rebassa-Mansergas}}]{gaensicke2006}
{G{\"a}nsicke}, B.~T., {Marsh}, T.~R., {Southworth}, J., \&
  {Rebassa-Mansergas}, A. 2006, Science, 314, 1908,
  \dodoi{10.1126/science.1135033}

\bibitem[{{Graham} {et~al.}(1990){Graham}, {Matthews}, {Neugebauer}, \&
  {Soifer}}]{graham1990}
{Graham}, J.~R., {Matthews}, K., {Neugebauer}, G., \& {Soifer}, B.~T. 1990,
  \apj, 357, 216, \dodoi{10.1086/168907}

\bibitem[{{Grether} \& {Lineweaver}(2006)}]{grether2006}
{Grether}, D., \& {Lineweaver}, C.~H. 2006, \apj, 640, 1051,
  \dodoi{10.1086/500161}

\bibitem[{{Hollands} {et~al.}(2018){Hollands}, {G{\"a}nsicke}, \&
  {Koester}}]{hollands2018a}
{Hollands}, M.~A., {G{\"a}nsicke}, B.~T., \& {Koester}, D. 2018, \mnras, 477,
  93, \dodoi{10.1093/mnras/sty592}

\bibitem[{{Jura}(2003)}]{jura2003}
{Jura}, M. 2003, \apjl, 584, L91, \dodoi{10.1086/374036}

\bibitem[{{Jura} {et~al.}(2007{\natexlab{a}}){Jura}, {Farihi}, \&
  {Zuckerman}}]{jura2007b}
{Jura}, M., {Farihi}, J., \& {Zuckerman}, B. 2007{\natexlab{a}}, \apj, 663,
  1285, \dodoi{10.1086/518767}

\bibitem[{{Jura} {et~al.}(2009){Jura}, {Farihi}, \& {Zuckerman}}]{jura2009a}
---. 2009, \aj, 137, 3191, \dodoi{10.1088/0004-6256/137/2/3191}

\bibitem[{{Jura} {et~al.}(2007{\natexlab{b}}){Jura}, {Farihi}, {Zuckerman}, \&
  {Becklin}}]{jura2007a}
{Jura}, M., {Farihi}, J., {Zuckerman}, B., \& {Becklin}, E.~E.
  2007{\natexlab{b}}, \aj, 133, 1927, \dodoi{10.1086/512734}

\bibitem[{{Keller} {et~al.}(2007){Keller}, {Schmidt}, {Bessell}, {Conroy},
  {Francis}, {Granlund}, {Kowald}, {Oates}, {Martin-Jones}, {Preston},
  {Tisserand}, {Vaccarella}, \& {Waterson}}]{keller2007}
{Keller}, S.~C., {Schmidt}, B.~P., {Bessell}, M.~S., {et~al.} 2007, \pasa, 24,
  1, \dodoi{10.1071/AS07001}

\bibitem[{{Kenyon} \& {Bromley}(2017{\natexlab{a}})}]{kenyon2017a}
{Kenyon}, S.~J., \& {Bromley}, B.~C. 2017{\natexlab{a}}, \apj, 844, 116,
  \dodoi{10.3847/1538-4357/aa7b85}

\bibitem[{{Kenyon} \& {Bromley}(2017{\natexlab{b}})}]{kenyon2017b}
---. 2017{\natexlab{b}}, \apj, 850, 50, \dodoi{10.3847/1538-4357/aa9570}

\bibitem[{{Klein} {et~al.}(2010){Klein}, {Jura}, {Koester}, {Zuckerman}, \&
  {Melis}}]{klein2010}
{Klein}, B., {Jura}, M., {Koester}, D., {Zuckerman}, B., \& {Melis}, C. 2010,
  \apj, 709, 950, \dodoi{10.1088/0004-637X/709/2/950}

\bibitem[{{Klein} {et~al.}(2021){Klein}, {Doyle}, {Zuckerman}, {Dufour},
  {Blouin}, {Melis}, {Weinberger}, \& {Young}}]{klein2021}
{Klein}, B.~L., {Doyle}, A.~E., {Zuckerman}, B., {et~al.} 2021, \apj, 914, 61,
  \dodoi{10.3847/1538-4357/abe40b}

\bibitem[{{Koester}(2010)}]{koester2010}
{Koester}, D. 2010, \memsai, 81, 921

\bibitem[{{Koester} {et~al.}(1997){Koester}, {Provencal}, \&
  {Shipman}}]{koester1997}
{Koester}, D., {Provencal}, J., \& {Shipman}, H.~L. 1997, \aap, 320, L57

\bibitem[{{Lai} {et~al.}(2021){Lai}, {Dennihy}, {Xu}, {Nitta}, {Kleinman},
  {Leggett}, {Bonsor}, {Hodgkin}, {Rebassa-Mansergas}, \& {Rogers}}]{lai2021}
{Lai}, S., {Dennihy}, E., {Xu}, S., {et~al.} 2021, \apj, 920, 156,
  \dodoi{10.3847/1538-4357/ac1354}

\bibitem[{{Lisse} {et~al.}(2009){Lisse}, {Chen}, {Wyatt}, {Morlok}, {Song},
  {Bryden}, \& {Sheehan}}]{lisse2009}
{Lisse}, C.~M., {Chen}, C.~H., {Wyatt}, M.~C., {et~al.} 2009, \apj, 701, 2019,
  \dodoi{10.1088/0004-637X/701/2/2019}

\bibitem[{{Lisse} {et~al.}(2006){Lisse}, {VanCleve}, {Adams}, {A'Hearn},
  {Fern{\'a}ndez}, {Farnham}, {Armus}, {Grillmair}, {Ingalls}, {Belton},
  {Groussin}, {McFadden}, {Meech}, {Schultz}, {Clark}, {Feaga}, \&
  {Sunshine}}]{lisse2006}
{Lisse}, C.~M., {VanCleve}, J., {Adams}, A.~C., {et~al.} 2006, Science, 313,
  635, \dodoi{10.1126/science.1124694}

\bibitem[{{Lisse} {et~al.}(2020){Lisse}, {Meng}, {Sitko}, {Morlok}, {Johnson},
  {Jackson}, {Vervack}, {Chen}, {Wolk}, {Lucas}, {Marengo}, \&
  {Britt}}]{lisse2020}
{Lisse}, C.~M., {Meng}, H.~Y.~A., {Sitko}, M.~L., {et~al.} 2020, \apj, 894,
  116, \dodoi{10.3847/1538-4357/ab7b80}

\bibitem[{{Malamud} {et~al.}(2021){Malamud}, {Grishin}, \&
  {Brouwers}}]{malamud2021}
{Malamud}, U., {Grishin}, E., \& {Brouwers}, M. 2021, \mnras, 501, 3806,
  \dodoi{10.1093/mnras/staa3940}

\bibitem[{{Malamud} \& {Perets}(2020)}]{malamud2020a}
{Malamud}, U., \& {Perets}, H.~B. 2020, \mnras, 492, 5561,
  \dodoi{10.1093/mnras/staa142}

\bibitem[{{Martin} {et~al.}(2005){Martin}, {Fanson}, {Schiminovich},
  {Morrissey}, {Friedman}, {Barlow}, {Conrow}, {Grange}, {Jelinsky},
  {Milliard}, {Siegmund}, {Bianchi}, {Byun}, {Donas}, {Forster}, {Heckman},
  {Lee}, {Madore}, {Malina}, {Neff}, {Rich}, {Small}, {Surber}, {Szalay},
  {Welsh}, \& {Wyder}}]{martin2005}
{Martin}, D.~C., {Fanson}, J., {Schiminovich}, D., {et~al.} 2005, \apjl, 619,
  L1, \dodoi{10.1086/426387}

\bibitem[{{Melis} \& {Dufour}(2017)}]{melis2017}
{Melis}, C., \& {Dufour}, P. 2017, \apj, 834, 1,
  \dodoi{10.3847/1538-4357/834/1/1}

\bibitem[{{Melis} {et~al.}(2010){Melis}, {Jura}, {Albert}, {Klein}, \&
  {Zuckerman}}]{melis2010}
{Melis}, C., {Jura}, M., {Albert}, L., {Klein}, B., \& {Zuckerman}, B. 2010,
  \apj, 722, 1078, \dodoi{10.1088/0004-637X/722/2/1078}

\bibitem[{{Melis} {et~al.}(2020){Melis}, {Klein}, {Doyle}, {Weinberger},
  {Zuckerman}, \& {Dufour}}]{melis2020}
{Melis}, C., {Klein}, B., {Doyle}, A.~E., {et~al.} 2020, \apj, 905, 56,
  \dodoi{10.3847/1538-4357/abbdfa}

\bibitem[{{Metzger} \& {Rafikov}(2012)}]{metzger2012}
{Metzger}, B.~D., \& {Rafikov}, Roman R.~{Bochkarev}, K.~V. 2012, \mnras, 423,
  505, \dodoi{10.1111/j.1365-2966.2012.20895.x}

\bibitem[{{Miles} {et~al.}(2023){Miles}, {Biller}, {Patapis}, {Worthen},
  {Rickman}, {Hoch}, {Skemer}, {Perrin}, {Whiteford}, {Chen}, {Sargent},
  {Mukherjee}, {Morley}, {Moran}, {Bonnefoy}, {Petrus}, {Carter}, {Choquet},
  {Hinkley}, {Ward-Duong}, {Leisenring}, {Millar-Blanchaer}, {Pueyo}, {Ray},
  {Sallum}, {Stapelfeldt}, {Stone}, {Wang}, {Absil}, {Balmer}, {Boccaletti},
  {Bonavita}, {Booth}, {Bowler}, {Chauvin}, {Christiaens}, {Currie},
  {Danielski}, {Fortney}, {Girard}, {Grady}, {Greenbaum}, {Henning}, {Hines},
  {Janson}, {Kalas}, {Kammerer}, {Kennedy}, {Kenworthy}, {Kervella}, {Lagage},
  {Lew}, {Liu}, {Macintosh}, {Marino}, {Marley}, {Marois}, {Matthews},
  {Matthews}, {Mawet}, {McElwain}, {Metchev}, {Meyer}, {Molliere}, {Pantin},
  {Quirrenbach}, {Rebollido}, {Ren}, {Schneider}, {Vasist}, {Wyatt}, {Zhou},
  {Briesemeister}, {Bryan}, {Calissendorff}, {Cantalloube}, {Cugno}, {De
  Furio}, {Dupuy}, {Factor}, {Faherty}, {Fitzgerald}, {Franson}, {Gonzales},
  {Hood}, {Howe}, {Kraus}, {Kuzuhara}, {Lagrange}, {Lawson}, {Lazzoni}, {Liu},
  {Llop-Sayson}, {Lloyd}, {Martinez}, {Mazoyer}, {Quanz}, {Redai}, {Samland},
  {Schlieder}, {Tamura}, {Tan}, {Uyama}, {Vigan}, {Vos}, {Wagner}, {Wolff},
  {Ygouf}, {Zhang}, {Zhang}, \& {Zhang}}]{miles2023}
{Miles}, B.~E., {Biller}, B.~A., {Patapis}, P., {et~al.} 2023, \apjl, 946, L6,
  \dodoi{10.3847/2041-8213/acb04a}

\bibitem[{{Mullally} {et~al.}(2007){Mullally}, {Kilic}, {Reach}, {Kuchner},
  {von Hippel}, {Burrows}, \& {Winget}}]{mullally2007}
{Mullally}, F., {Kilic}, M., {Reach}, W.~T., {et~al.} 2007, \apjs, 171, 206,
  \dodoi{10.1086/511858}

\bibitem[{{Noor} {et~al.}(2024){Noor}, {Farihi}, {Hollands}, \&
  {Toonen}}]{noor2024}
{Noor}, H.~T., {Farihi}, J., {Hollands}, M., \& {Toonen}, S. 2024, \mnras, 529,
  2910, \dodoi{10.1093/mnras/stae731}

\bibitem[{{Probst}(1983)}]{probst1983}
{Probst}, R.~G. 1983, \apjs, 53, 335, \dodoi{10.1086/190893}

\bibitem[{{Rafikov}(2011)}]{rafikov2011a}
{Rafikov}, R.~R. 2011, \apjl, 732, L3, \dodoi{10.1088/2041-8205/732/1/L3}

\bibitem[{{Rafikov} \& {Garmilla}(2012)}]{rafikov2012}
{Rafikov}, R.~R., \& {Garmilla}, J.~A. 2012, \apj, 760, 123,
  \dodoi{10.1088/0004-637X/760/2/123}

\bibitem[{{Reach} {et~al.}(2005){Reach}, {Kuchner}, {von Hippel}, {Burrows},
  {Mullally}, {Kilic}, \& {Winget}}]{reach2005b}
{Reach}, W.~T., {Kuchner}, M.~J., {von Hippel}, T., {et~al.} 2005, \apjl, 635,
  L161, \dodoi{10.1086/499561}

\bibitem[{{Reach} {et~al.}(2009){Reach}, {Lisse}, {von Hippel}, \&
  {Mullally}}]{reach2009}
{Reach}, W.~T., {Lisse}, C., {von Hippel}, T., \& {Mullally}, F. 2009, \apj,
  693, 697, \dodoi{10.1088/0004-637X/693/1/697}

\bibitem[{{Rieke} {et~al.}(2015){Rieke}, {Wright}, {B{\"o}ker}, {Bouwman},
  {Colina}, {Glasse}, {Gordon}, {Greene}, {G{\"u}del}, {Henning}, {Justtanont},
  {Lagage}, {Meixner}, {N{\o}rgaard-Nielsen}, {Ray}, {Ressler}, {van Dishoeck},
  \& {Waelkens}}]{rieke2015}
{Rieke}, G.~H., {Wright}, G.~S., {B{\"o}ker}, T., {et~al.} 2015, \pasp, 127,
  584, \dodoi{10.1086/682252}

\bibitem[{{Rocchetto} {et~al.}(2015){Rocchetto}, {Farihi}, \&
  {G{\"a}nsicke}}]{rocchetto2015}
{Rocchetto}, M., {Farihi}, J., \& {G{\"a}nsicke}, B.~T.~{Bergfors}, C. 2015,
  \mnras, 449, 574, \dodoi{10.1093/mnras/stv282}

\bibitem[{{Schatzman}(1945)}]{schatzman1945}
{Schatzman}, E. 1945, Annales d'Astrophysique, 8, 143

\bibitem[{{Skrutskie} {et~al.}(2006){Skrutskie}, {Cutri}, {Stiening},
  {Weinberg}, {Schneider}, {Carpenter}, {Beichman}, {Capps}, {Chester},
  {Elias}, {Huchra}, {Liebert}, {Lonsdale}, {Monet}, {Price}, {Seitzer},
  {Jarrett}, {Kirkpatrick}, {Gizis}, {Howard}, {Evans}, {Fowler}, {Fullmer},
  {Hurt}, {Light}, {Kopan}, {Marsh}, {McCallon}, {Tam}, {Van Dyk}, \&
  {Wheelock}}]{skrutskie2006}
{Skrutskie}, M.~F., {Cutri}, R.~M., {Stiening}, R., {et~al.} 2006, \aj, 131,
  1163, \dodoi{10.1086/498708}

\bibitem[{{Smallwood} {et~al.}(2018){Smallwood}, {Martin}, {Livio}, \&
  {Lubow}}]{smallwood2018}
{Smallwood}, J.~L., {Martin}, R.~G., {Livio}, M., \& {Lubow}, S.~H. 2018,
  \mnras, 480, 57, \dodoi{10.1093/mnras/sty1819}

\bibitem[{{Song} {et~al.}(2005){Song}, {Zuckerman}, {Weinberger}, \&
  {Becklin}}]{song2005}
{Song}, I., {Zuckerman}, B., {Weinberger}, A.~J., \& {Becklin}, E.~E. 2005,
  \nat, 436, 363, \dodoi{10.1038/nature03853}

\bibitem[{{Steckloff} {et~al.}(2021){Steckloff}, {Debes}, {Steele}, {Johnson},
  {Adams}, {Jacobson}, \& {Springmann}}]{steckloff2021}
{Steckloff}, J.~K., {Debes}, J., {Steele}, A., {et~al.} 2021, \apjl, 913, L31,
  \dodoi{10.3847/2041-8213/abfd39}

\bibitem[{{Swan} {et~al.}(2023){Swan}, {Farihi}, {Melis}, {Dufour}, {Desch},
  {Koester}, \& {Guo}}]{swan2023}
{Swan}, A., {Farihi}, J., {Melis}, C., {et~al.} 2023, \mnras, 526, 3815,
  \dodoi{10.1093/mnras/stad2867}

\bibitem[{{Swan} {et~al.}(2021){Swan}, {Kenyon}, {Farihi}, {Dennihy},
  {G{\"a}nsicke}, {Hermes}, {Melis}, \& {von Hippel}}]{swan2021}
{Swan}, A., {Kenyon}, S.~J., {Farihi}, J., {et~al.} 2021, \mnras, 506, 432,
  \dodoi{10.1093/mnras/stab1738}

\bibitem[{{Tonry} {et~al.}(2012){Tonry}, {Stubbs}, {Lykke}, {Doherty},
  {Shivvers}, {Burgett}, {Chambers}, {Hodapp}, {Kaiser}, {Kudritzki},
  {Magnier}, {Morgan}, {Price}, \& {Wainscoat}}]{tonry2012}
{Tonry}, J.~L., {Stubbs}, C.~W., {Lykke}, K.~R., {et~al.} 2012, \apj, 750, 99,
  \dodoi{10.1088/0004-637X/750/2/99}

\bibitem[{{Weinberger} {et~al.}(2011){Weinberger}, {Becklin}, {Song}, \&
  {Zuckerman}}]{weinberger2011}
{Weinberger}, A.~J., {Becklin}, E.~E., {Song}, I., \& {Zuckerman}, B. 2011,
  \apj, 726, 72, \dodoi{10.1088/0004-637X/726/2/72}

\bibitem[{{Williams} {et~al.}(2024){Williams}, {Gaensicke}, {Swan}, {O'Brien},
  {Izquierdo}, {Cutolo}, \& {Cunningham}}]{williams2024}
{Williams}, J., {Gaensicke}, B., {Swan}, A., {et~al.} 2024, arXiv e-prints,
  arXiv:2409.16046, \dodoi{10.48550/arXiv.2409.16046}

\bibitem[{{Wilson} {et~al.}(2016){Wilson}, {G{\"a}nsicke}, \&
  {Farihi}}]{wilsond2016}
{Wilson}, D.~J., {G{\"a}nsicke}, B.~T., \& {Farihi}, Jay~{Koester}, D. 2016,
  \mnras, 459, 3282, \dodoi{10.1093/mnras/stw844}

\bibitem[{{Wright} {et~al.}(2010){Wright}, {Eisenhardt}, {Mainzer}, {Ressler},
  {Cutri}, {Jarrett}, {Kirkpatrick}, {Padgett}, {McMillan}, {Skrutskie},
  {Stanford}, {Cohen}, {Walker}, {Mather}, {Leisawitz}, {Gautier}, {McLean},
  {Benford}, {Lonsdale}, {Blain}, {Mendez}, {Irace}, {Duval}, {Liu}, {Royer},
  {Heinrichsen}, {Howard}, {Shannon}, {Kendall}, {Walsh}, {Larsen}, {Cardon},
  {Schick}, {Schwalm}, {Abid}, {Fabinsky}, {Naes}, \& {Tsai}}]{wright2010}
{Wright}, E.~L., {Eisenhardt}, P. R.~M., {Mainzer}, A.~K., {et~al.} 2010, \aj,
  140, 1868, \dodoi{10.1088/0004-6256/140/6/1868}

\bibitem[{{Xu} {et~al.}(2017){Xu}, {Zuckerman}, {Dufour}, {Young}, {Klein}, \&
  {Jura}}]{xu2017}
{Xu}, S., {Zuckerman}, B., {Dufour}, P., {et~al.} 2017, \apjl, 836, L7,
  \dodoi{10.3847/2041-8213/836/1/L7}

\bibitem[{{Zuckerman} \& {Becklin}(1987)}]{zuckerman1987}
{Zuckerman}, B., \& {Becklin}, E.~E. 1987, \nat, 330, 138,
  \dodoi{10.1038/330138a0}

\bibitem[{{Zuckerman} {et~al.}(2011){Zuckerman}, {Koester}, {Dufour}, {Melis},
  {Klein}, \& {Jura}}]{zuckerman2011}
{Zuckerman}, B., {Koester}, D., {Dufour}, P., {et~al.} 2011, \apj, 739, 101,
  \dodoi{10.1088/0004-637X/739/2/101}

\bibitem[{{Zuckerman} {et~al.}(2007){Zuckerman}, {Koester}, {Melis}, {Hansen},
  \& {Jura}}]{zuckerman2007}
{Zuckerman}, B., {Koester}, D., {Melis}, C., {Hansen}, B.~M., \& {Jura}, M.
  2007, \apj, 671, 872, \dodoi{10.1086/522223}

\bibitem[{{Zuckerman} {et~al.}(2003){Zuckerman}, {Koester}, {Reid}, \&
  {H{\"u}nsch}}]{zuckerman2003}
{Zuckerman}, B., {Koester}, D., {Reid}, I.~N., \& {H{\"u}nsch}, M. 2003, \apj,
  596, 477, \dodoi{10.1086/377492}

\end{thebibliography}
\bibliographystyle{aasjournal}

\appendix

% FIGURE %%%
\begin{figure*}[h!]
\includegraphics[width=0.33\textwidth]{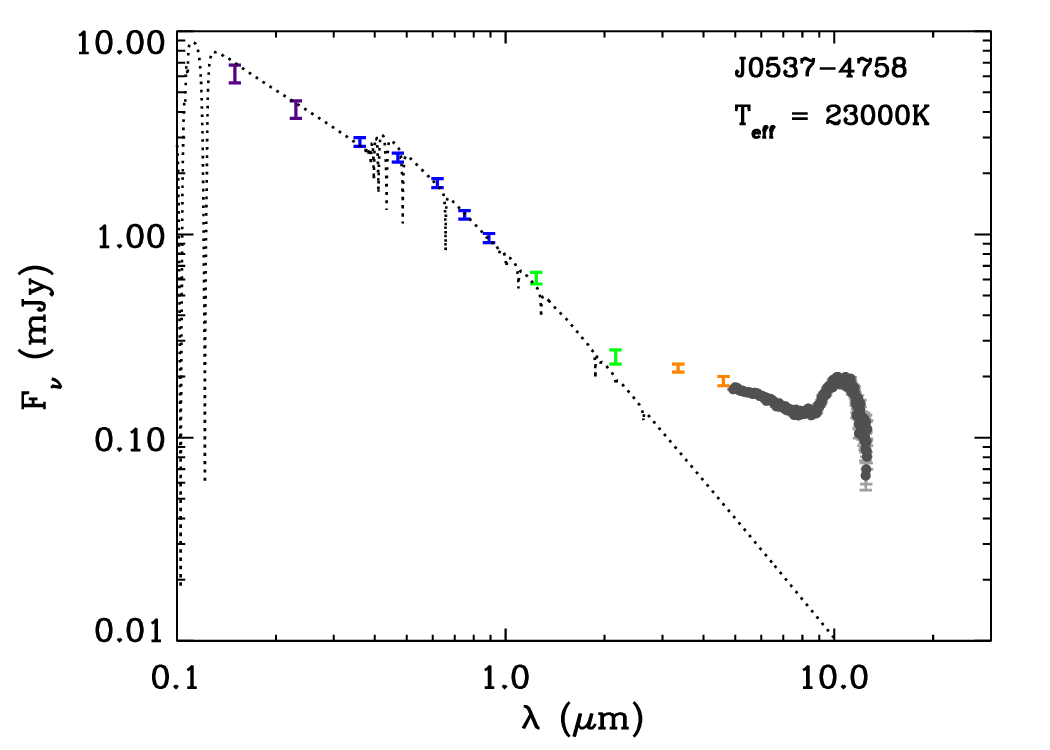}
\includegraphics[width=0.33\textwidth]{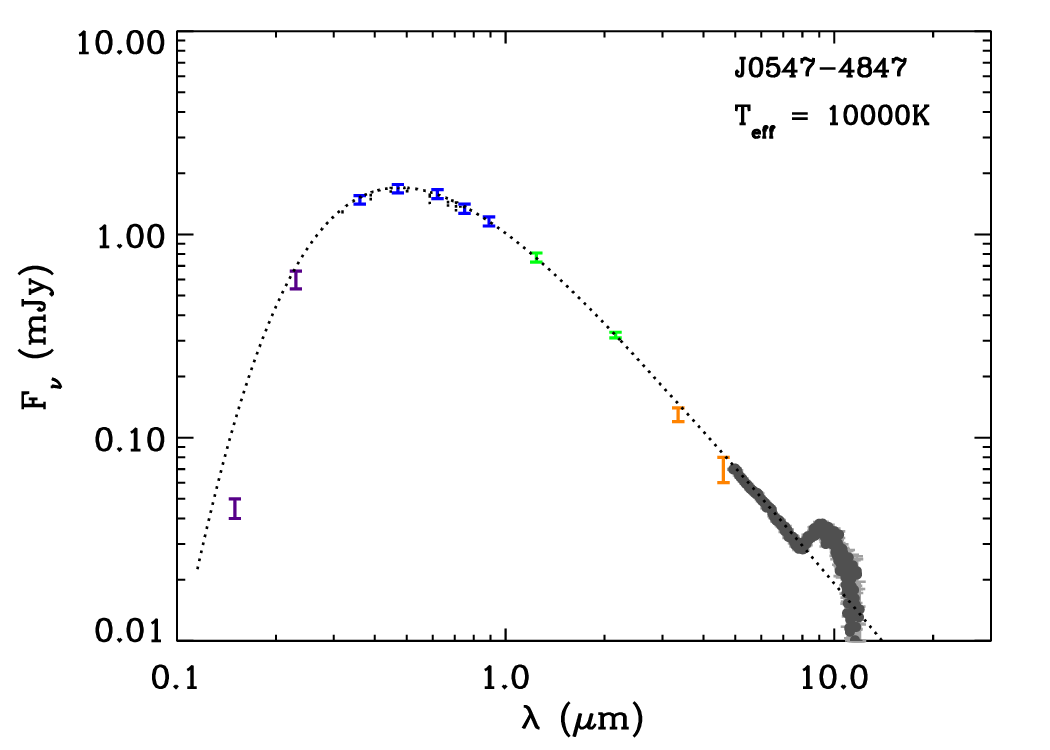}
\includegraphics[width=0.33\textwidth]{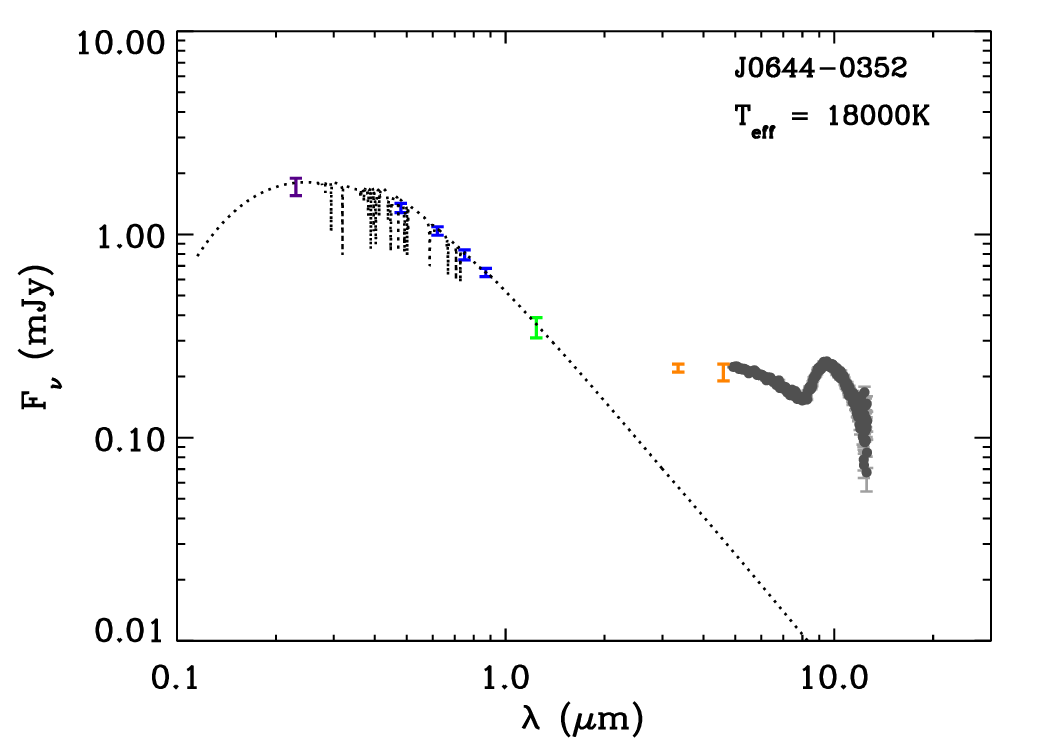}
\includegraphics[width=0.33\textwidth]{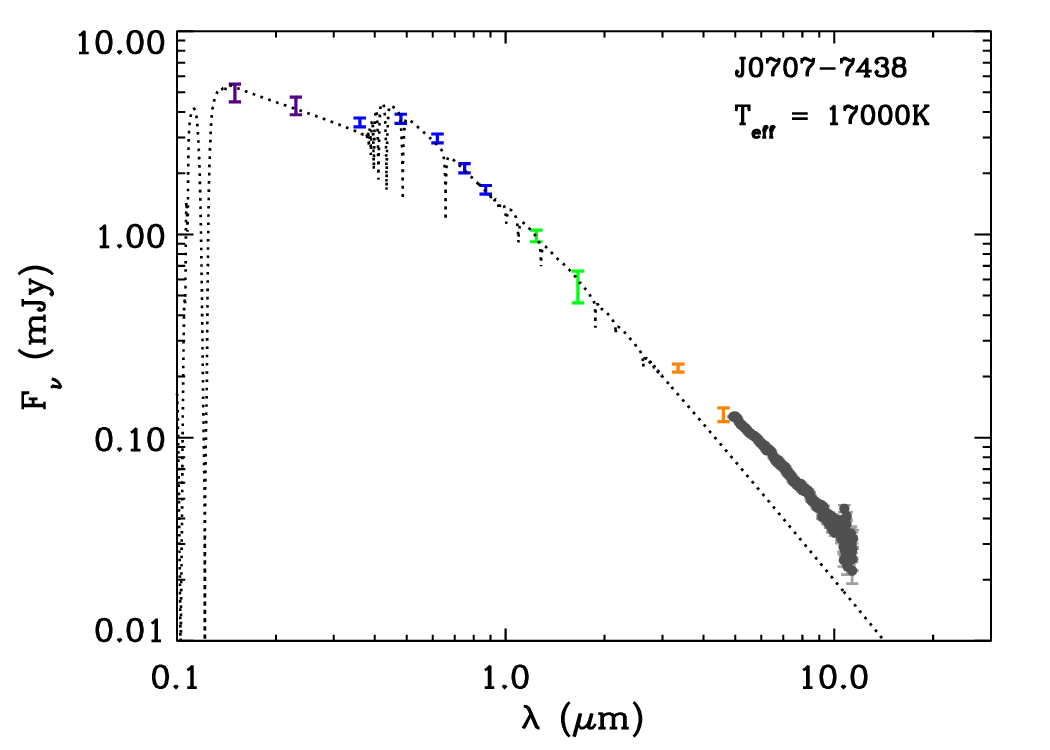}
\includegraphics[width=0.33\textwidth]{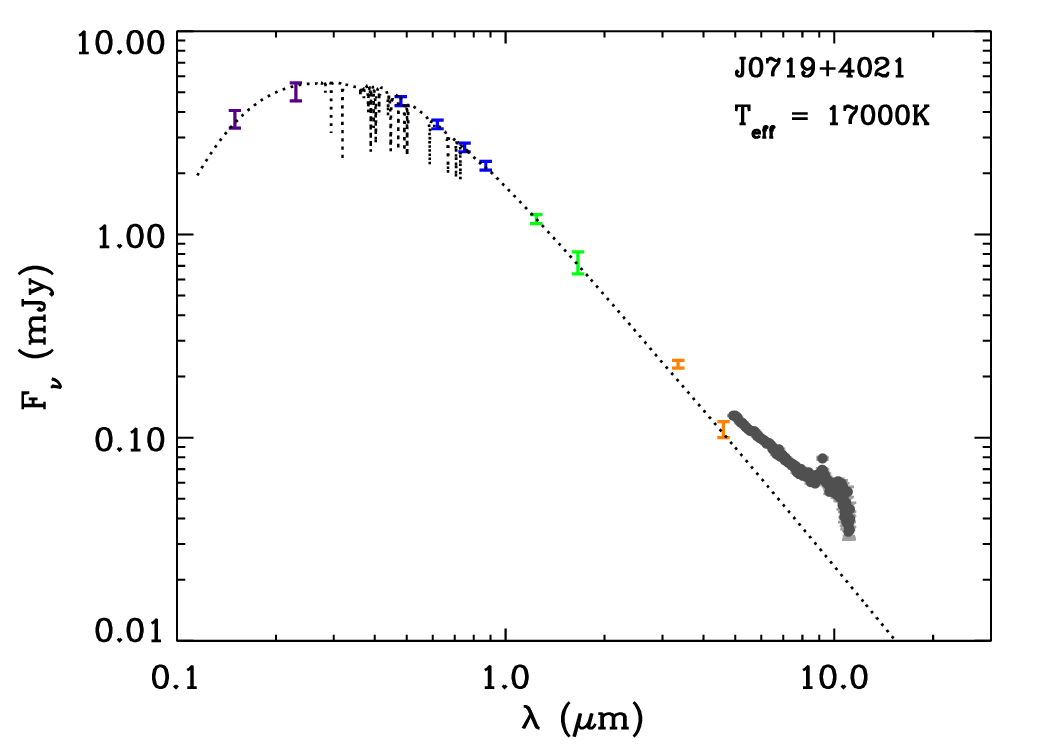}
\includegraphics[width=0.33\textwidth]{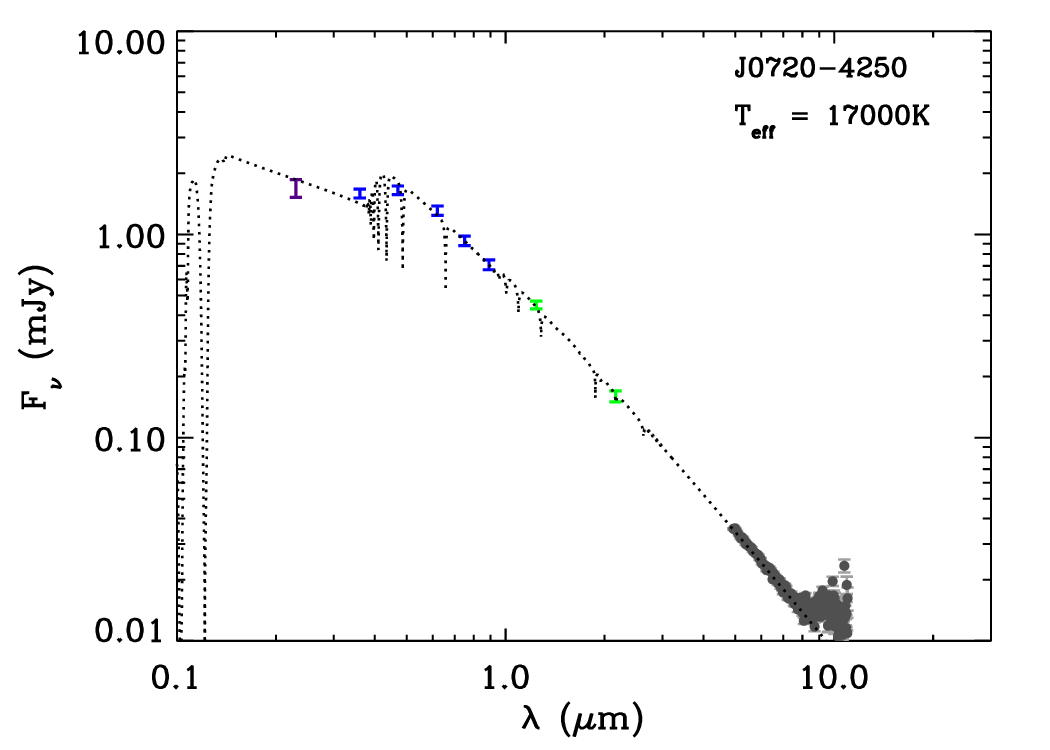}
\includegraphics[width=0.33\textwidth]{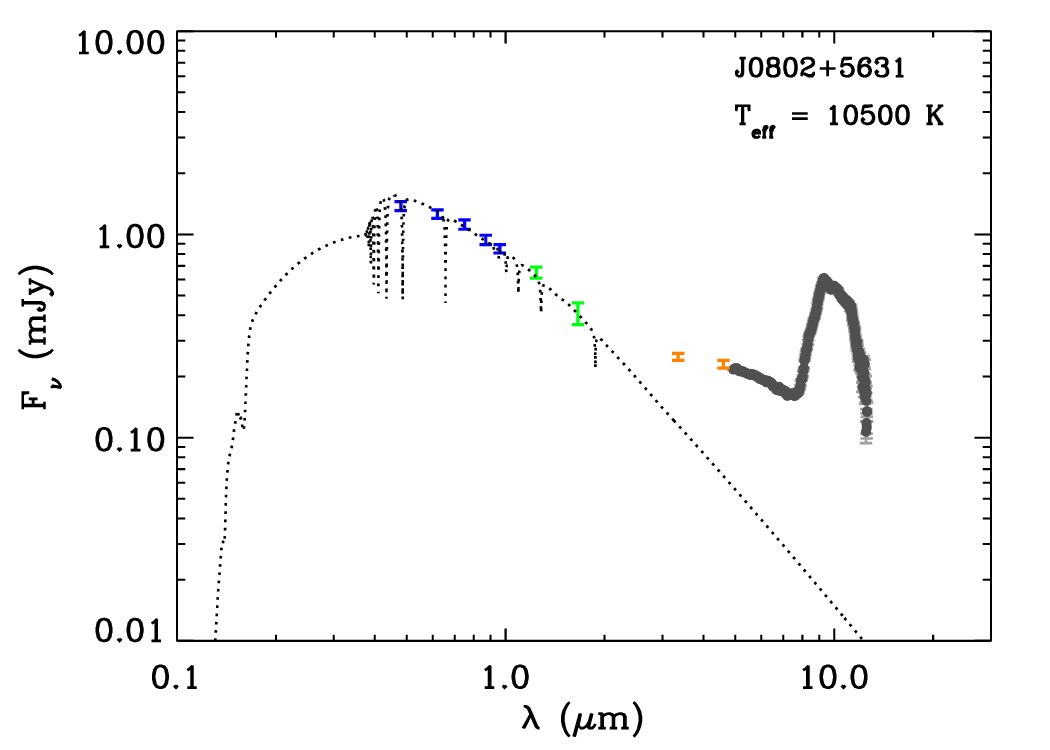}
\includegraphics[width=0.33\textwidth]{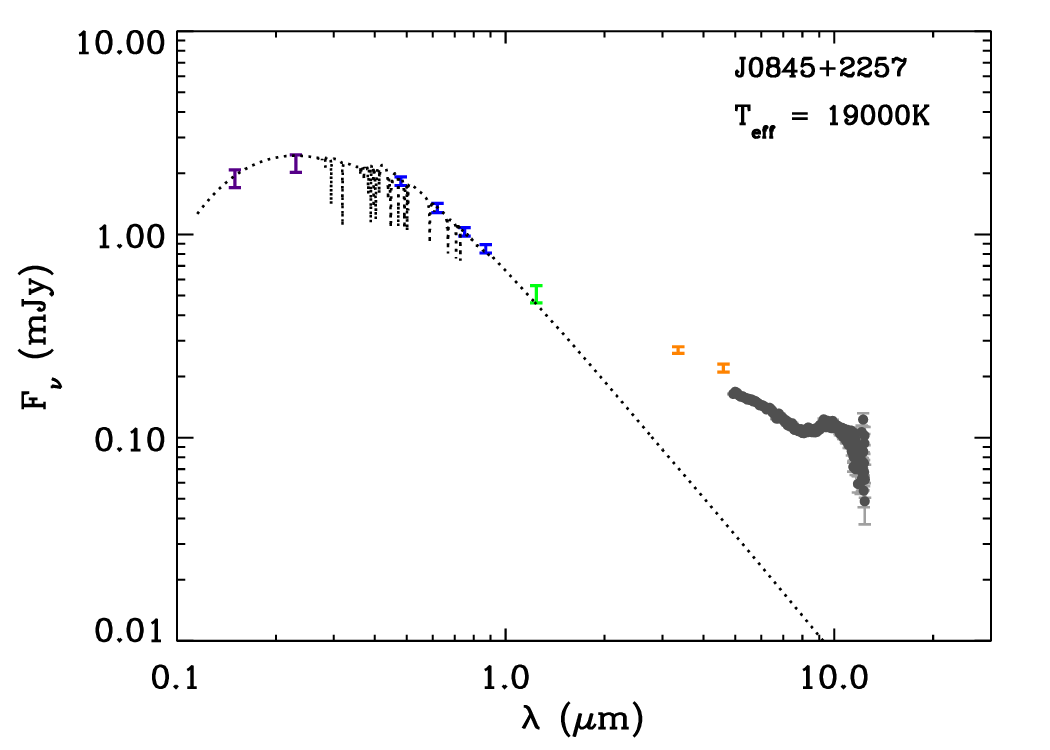}
\includegraphics[width=0.33\textwidth]{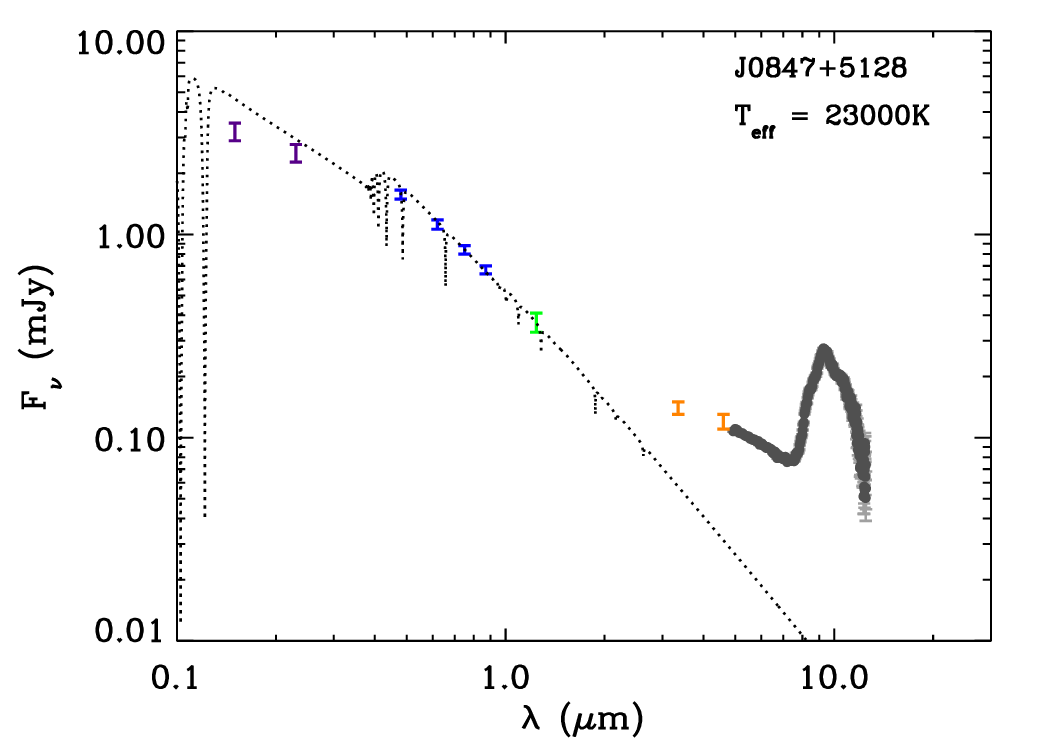}
\includegraphics[width=0.33\textwidth]{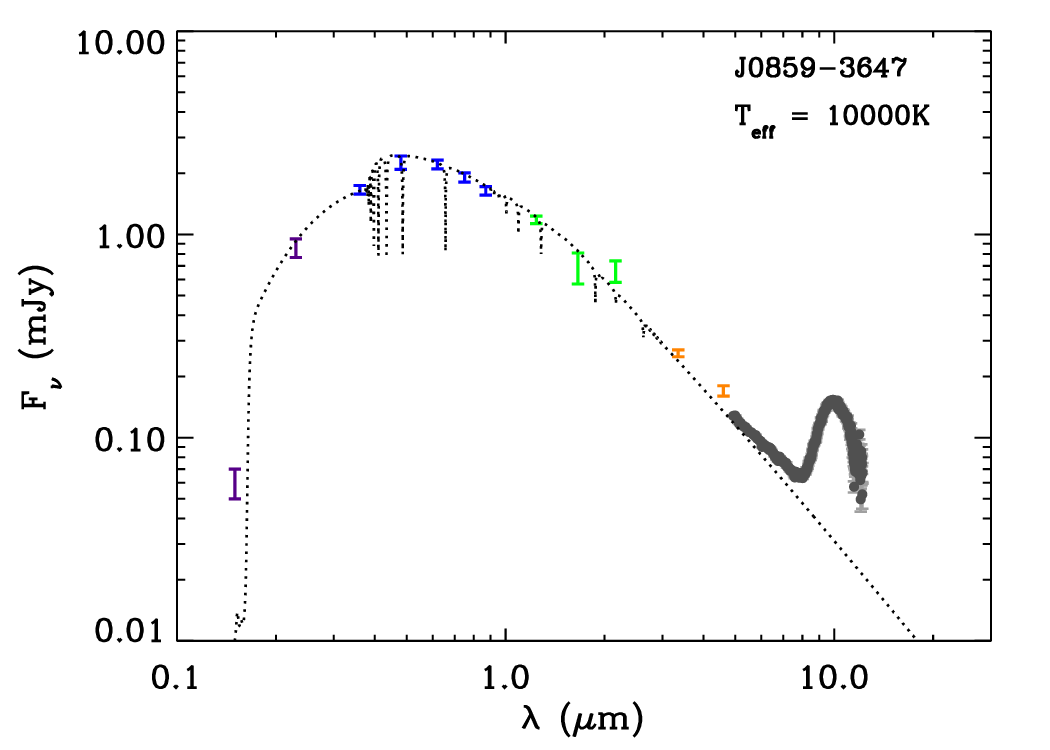}
\includegraphics[width=0.33\textwidth]{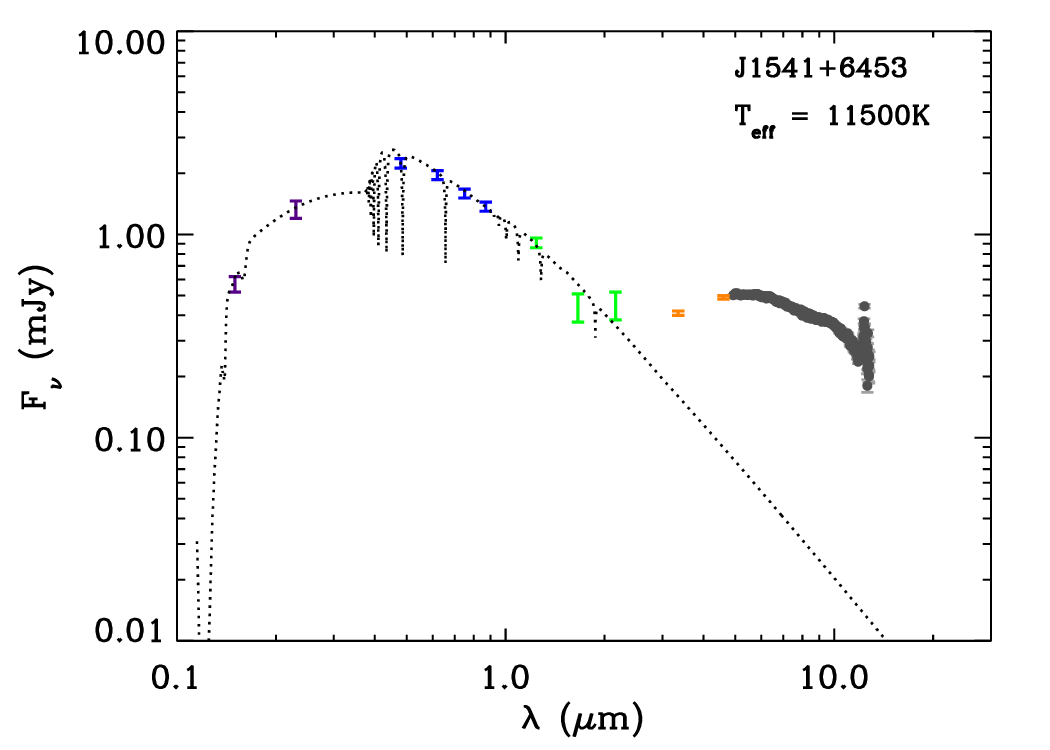}
\includegraphics[width=0.33\textwidth]{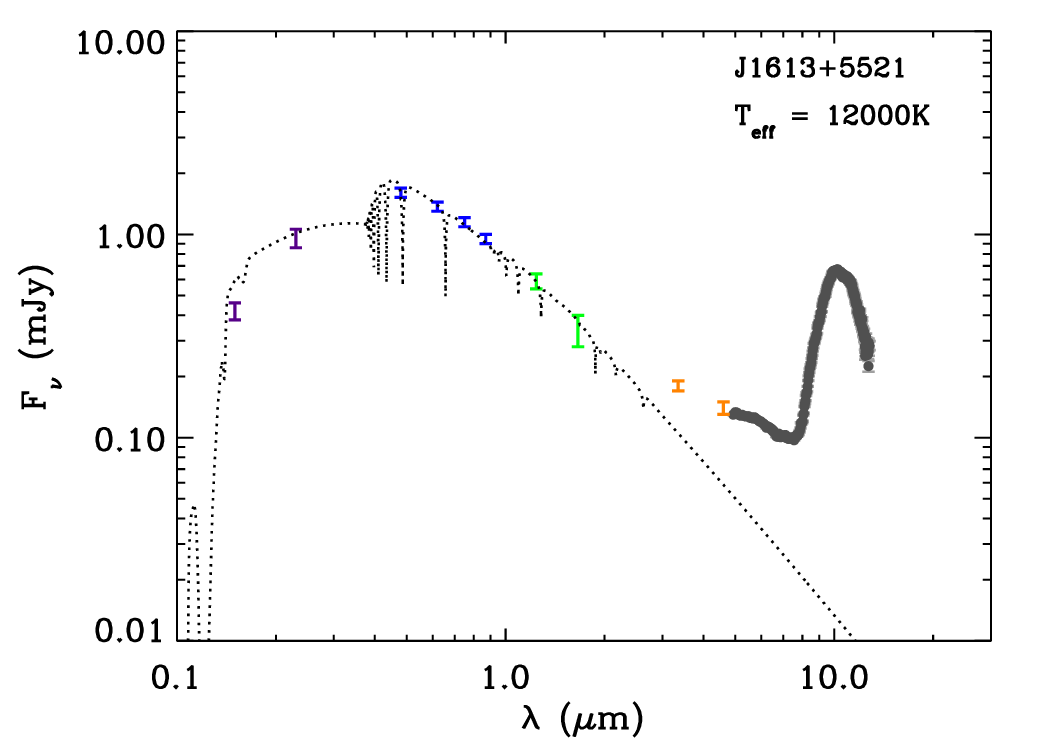}
\vskip 0pt
\caption{The full photometric spectral energy distributions of the 12 stars with infrared excess plotted in Figure~\ref{fig_main}.  Photometric data are shown as error bars and are color-coded as follows; purple is {\em GALEX}, blue is Pan-STARRS, SDSS, or SkyMapper, green is 2MASS or VISTA, and orange is {\em WISE}.  The fitting process is described briefly in Section~2.  The MIRI LRS spectra are shown as grey data points with error bars, and have been truncated near or just beyond $12\,\upmu$m for cosmetic reasons owing to noise at these longer wavelengths.
\label{fig_app}}
\end{figure*}

\end{document}